\documentclass[reprint,preprintnumbers,onecolumn]{revtex4}

\usepackage{amsthm}
\usepackage{graphicx}
\usepackage{dcolumn}
\usepackage{bm}
\usepackage{amssymb}
\usepackage{verbatim}
\usepackage{amscd}
\usepackage{amssymb}
\usepackage{amsmath,amsfonts}
\usepackage{amsthm}
\usepackage{enumerate}
\newcommand{\tb}[1]{\textbf{#1}}

\theoremstyle{plain}
\newtheorem{theorem}{Theorem}
\theoremstyle{plain}
\newtheorem{lemma}{Lemma}
\theoremstyle{plain}
 
\theoremstyle{plain}

\theoremstyle{remark}

\theoremstyle{conjecture}

\theoremstyle{observation}

\theoremstyle{definition}

\theoremstyle{corollary}
\newtheorem{corollary}{Corollary}
\theoremstyle{definition}

\theoremstyle{definition}

\theoremstyle{assumption}

\theoremstyle{definition}

\theoremstyle{problem}

\theoremstyle{fact}
\newtheorem{fact}{Fact}

\begin{document}

\preprint{MIT-CTP 4326}

\title{Information storage capacity of discrete spin systems}
\author{Beni Yoshida}
\affiliation{Center for Theoretical Physics, Massachusetts Institute of Technology, Cambridge, Massachusetts 02139, USA}
\affiliation{Institute for Quantum Information and Matter, California Institute of Technology, Pasadena, California 91125, USA}

\date{\today}

\begin{abstract}
Understanding the limits imposed on information storage capacity of physical systems is a problem of fundamental and practical importance which bridges physics and information science. There is a well-known upper bound on the amount of information that can be stored reliably in a given volume of discrete spin systems which are supported by gapped local Hamiltonians. However, all the previously known systems were far below this theoretical bound, and it remained open whether there exists a gapped spin system that saturates this bound. Here, we present a construction of spin systems which saturate this theoretical limit asymptotically by borrowing an idea from fractal properties arising in the Sierpinski triangle. Our construction provides not only the best classical error-correcting code which is physically realizable as the energy ground space of gapped frustration-free Hamiltonians, but also a new research avenue for correlated spin phases with fractal spin configurations. 
\end{abstract}

\maketitle

\section{Introduction and summary of results}

Understanding the limits imposed on information storage capacity of physical systems is a problem of fundamental and practical importance which bridges physics and information science~\cite{Landauer88}. This problem has been addressed for continuum systems by Bekenstein~\cite{Bekenstein81}. He showed that it is not possible to store an infinite amount of information on a finite system and derived the well-celebrated bound on the number of logical bits that can be stored inside a finite region:
\begin{align}
S \leq \frac{2 \pi k_{B} LE}{\hbar c}
\end{align}
where $S$ is the amount of information stored, $L$ is the linear length of the region, and $E$ is the total energy. While the Bekenstein bound itself can be derived from simple quantum mechanical calculations, the most beautiful outcome concerning the Bekenstein bound is that black holes saturate this theoretical limit, giving rise to the area law of black hole entropies~\cite{Hawking75}. This is essentially due to the observation that an object with a large amount of information (entropy) tends to have high energy, and will eventually turn into a black hole once its energy exceeds a critical value. This surprising connection between information theory and black hole physics is at the heart of the thermodynamic treatment of black holes and the holographic principle~\cite{Susskind95}. 

\begin{figure}[htb!]
\centering
\includegraphics[width=0.60\linewidth]{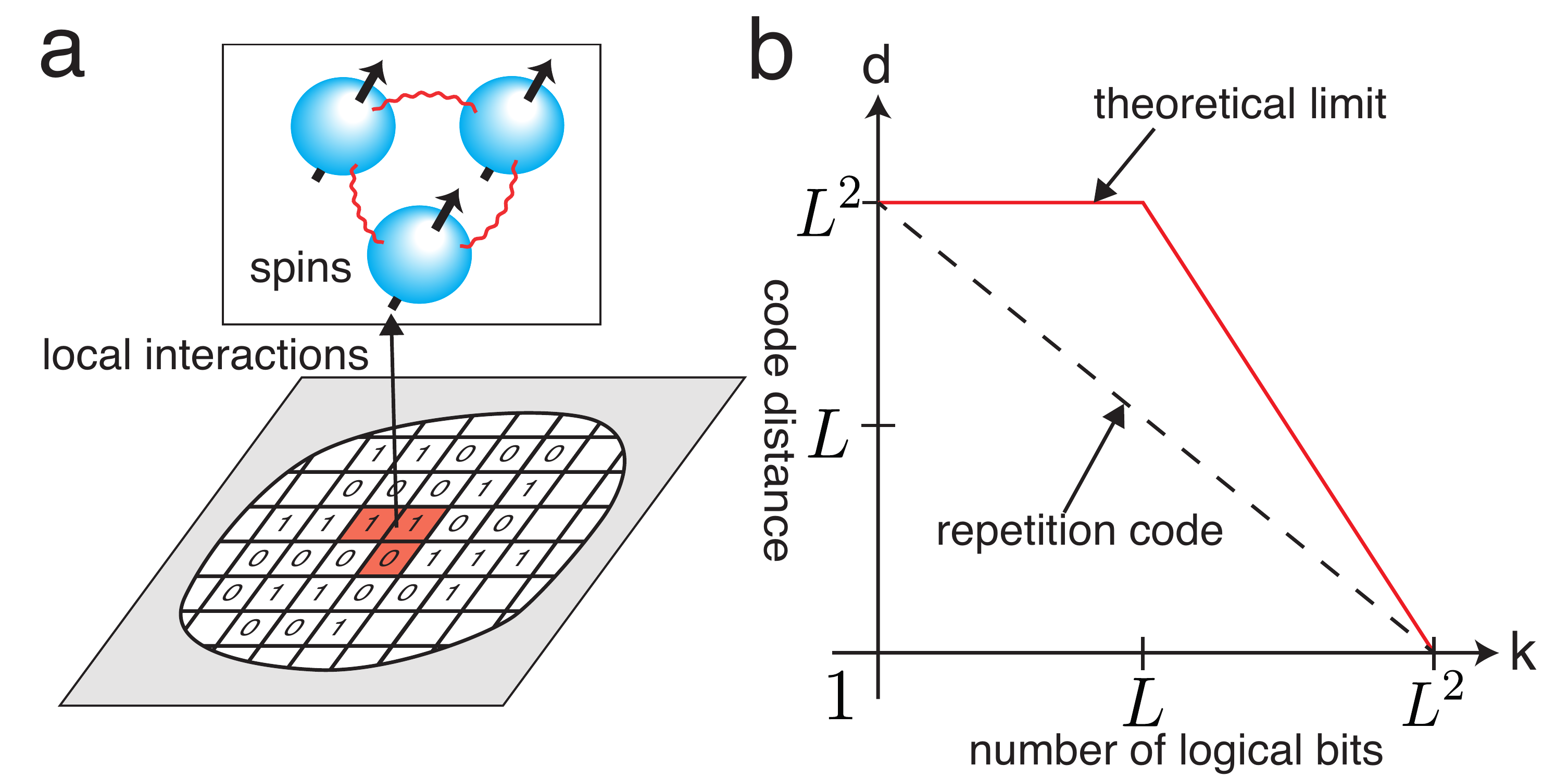}
\caption{(a) Storage of information in discrete spin systems via local interactions. (b) A theoretical upper bound on information storage capacity for $D=2$. The graph is shown in a logarithmic scale. The dotted line represents a family of repetition codes. 
} 
\label{fig_bound_graph}
\end{figure}

Recently, a similar question on information storage capacity for discrete spin systems on a lattice has been addressed. Consider discrete spin systems defined on a $D$-dimensional lattice which is governed by a local gapped Hamiltonian where $D$ is the spatial dimension. To be concrete, we consider commuting frustration-free Hamiltonians with local interaction terms, which are referred to as \emph{local codes}. Now, we think of encoding bits of information into degenerate gapped ground states of local codes. Then, the following bound is known to hold~\cite{Bravyi10}:
\begin{align}
kd^{1/D} \leq O(n)\label{eq:bound} 
\end{align} 
where $k$ is the number of encoded logical bits, $d$ is the code distance, and $n$ is the total number of spins when the energy ground space of a local Hamiltonian is viewed as the codeword space of an error-correcting code. A proportional factor on the right-hand side of the bound depends on the range of interaction terms. Note that the code distance $d$ is a quantitative measure of the reliability of encoded bits against errors. Thus, the bound above reveals a fundamental tradeoff between the amount of encoded information $k$ and the reliability of encoding $d$.

Motivated by a tremendous success of the Bekenstein bound and its significant impact on black hole physics, one may be naturally led to an analogous question on information storage capacity of discrete spin systems, concerning local codes which saturate the local code bound. This is a problem of practical importance since, in principle, such a local code would be the best error-correcting code that is physically realizable with frustration-free local Hamiltonians. This problem may also be of fundamental importance since such a local code may be viewed as an analog of a black hole for discrete spin systems in some appropriate interpretation which is yet to be discovered, and may be useful in further establishing the connection between continuum and discrete descriptions of space-time and quantum gravity~\cite{Penrose71, Rovelli95}. Finally, such a local code may be a candidate model of novel spin phases with exotic correlations which are beyond descriptions of known effective theories with mass gap, such as topological field theory.

However, finding a local code which saturates the bound turned out to be a challenging problem. In particular, previously found local codes were far below the bound as seen in Fig.~\ref{fig_bound_graph}(b). To gain some insights on the problem, let us look at a prototypical example of local codes on a two-dimensional lattice ($D=2$). A repetition code encodes $0$ and $1$ into repetitions of zeros and ones; $000\cdots$ and $111\cdots$, and can be physically realized as a local code through local ferromagnetic interactions. Since it encodes a single bit of information, it has $k=1$. A repetition code is known to be robust against errors since the originally encoded bit of information can be faithfully recovered provided that the number of damaged spins is less than $\frac{n}{2}$. A natural measure of the reliability of encoding is the number of different spin values in two codewords, which is called the Hamming distance in the coding theory language. Since all the spin values of codewords are different in a repetition code, the Hamming distance between codewords is $n$, and the code distance is $d=n$. Now, let us analyze coding properties of a repetition code in terms of the local code bound. For $D=2$, the local code bound is $k\sqrt{d}\leq O(n)$, and the repetition code is far below the theoretical limit. One may modify a repetition code by splitting the entire lattice into smaller subparts and using them as individual repetition codes. However, such a construction gives a family of local codes with $kd=n$ as shown with a dotted line in Fig.~\ref{fig_bound_graph}(b), which is still below the bound. There had been no local code with provably better coding properties than a family of repetition codes. Also, it should be noted that commercial memory devices, such as hard disc drives (HDD), are constructed with ferromagnetic materials which are physical realizations of repetition codes.\\

\tb{Main result:}
In this paper, we present a construction of local codes, called \emph{fractal codes}, which saturate the theoretical limit asymptotically:
\begin{align*}
k \sim O(L^{D-1}),\qquad d \sim O(L^{D-\epsilon})  
\end{align*}
for $D\geq 2$ where $\epsilon$ is an arbitrary small positive number, $L$ is the linear length of the lattice and $n=O(L^{D})$. \\

\begin{figure}[htb!]
\centering
\includegraphics[width=0.75\linewidth]{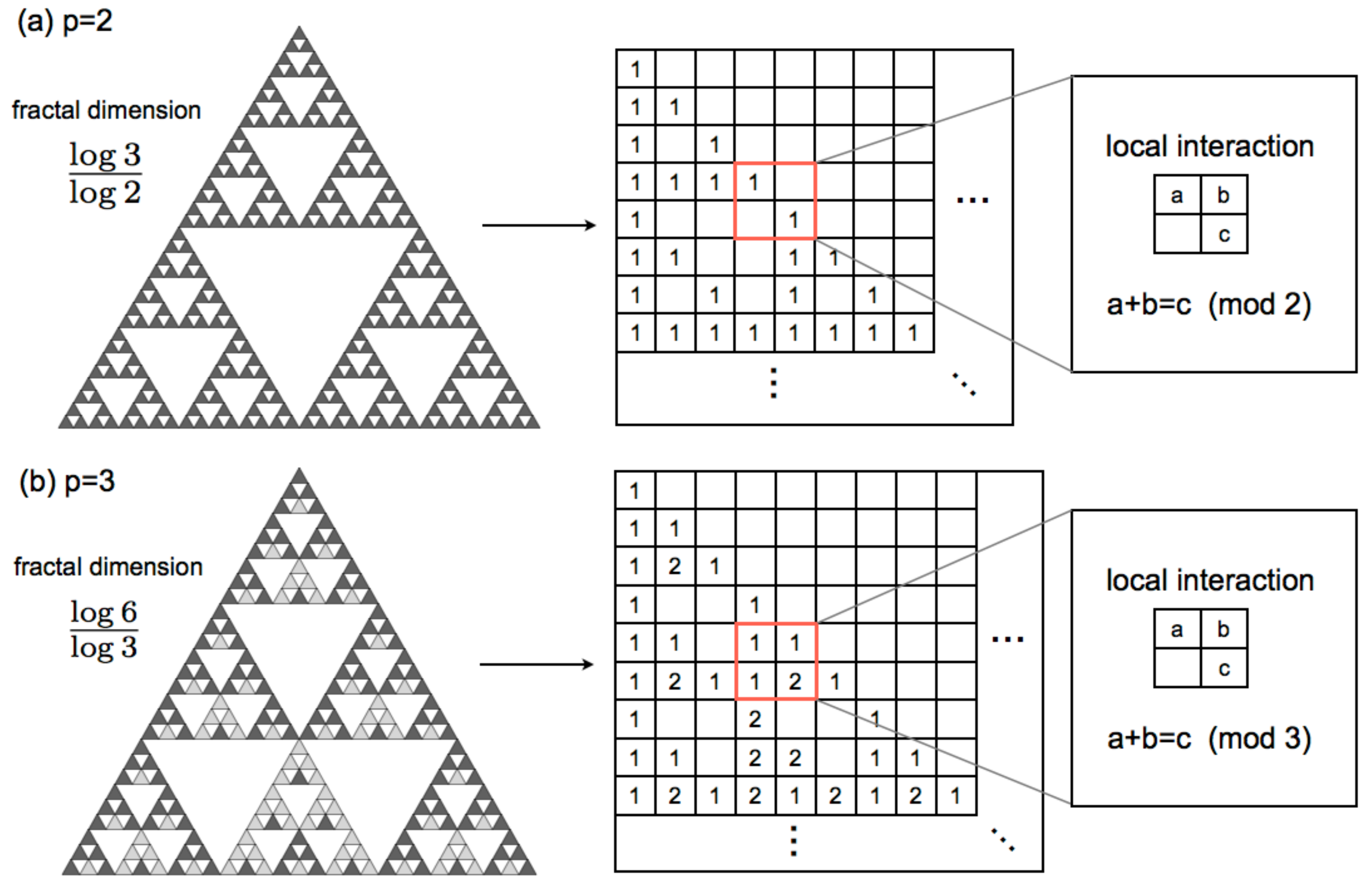}
\caption{Fractal codes. (a) The Sierpinski triangle and its physical realization on a square lattice ($p=2$). Filled elements are mapped to $1$s while unfilled elements are mapped to $0$s. Interaction terms are three-body. (b) A generalization of the Sierpinski triangle ($p=3$). Black elements are mapped to $1$s, grey elements are mapped to $2$s, and unfilled elements are mapped to $0$s.
} 
\label{fig_Sier}
\end{figure}

\tb{Fractal geometry as a code:}
Our construction borrows an idea from the Sierpinski triangle, a well-known example of fractal geometries. The Sierpinski triangle has self-similar properties where the same patterns appear repeatedly at different length scales (Fig.~\ref{fig_Sier}a). This peculiar geometric nature of the triangle is reflected in its non-integer dimensionality where the number of filled elements $L^{\log 3/ \log 2}$ grows as if the spatial dimension is $\frac{\log 3}{\log 2}\sim 1.585$. While the Sierpinski triangle had been long thought to be a mathematical object, it turned out that the triangle is physically realizable. Fig~\ref{fig_Sier}(a) shows a physical realization of the Sierpinski triangle on a square lattice via three-body interactions where each term is minimized when local constraints $c=a+b$ (mod $2$) on three neighboring spins are satisfied~\cite{Newman99}. It has been pointed out that such a fractal system, generated by cellular automaton, may be useful as an error-correcting code with an efficient decoder~\cite{Chowdhury94}. Recently, its coding properties have been predicted as~\cite{Bravyi10}:
\begin{align*}
k \sim O(L),\qquad d \sim O(L^{\frac{\log 3}{\log 2}})
\end{align*} 
based on numerical simulations along with analytical arguments for infinite lattices.

Despite a remarkable idea of constructing a local code based on Sierpinski triangle, previous works have two serious drawbacks. First, this fractal code is still far below the theoretical limit as seen in Fig.~\ref{fig_bound_graph}(b). Second, in order to prove the prediction of $d \sim O(L^{\frac{\log 3}{\log 2}})$, one needs to analyze Hamming distances between all the $2^{L}$ ground states and find the minimal Hamming distance, which is a formidable challenge both from analytical and computational perspectives.

We start by presenting the resolution of the first challenge. Our construction of fractal codes utilizes a generalization of Sierpinski triangle with higher-dimensional spins. To begin with, let us discuss fractal properties of Sierpinski triangle with three-dimensional spins where possible spin values are $0,1,2$ as shown in Fig.~\ref{fig_Sier}(b). The number of non-zero spins in this generalized Sierpinski triangle is $L^{\frac{\log 6}{\log 3}}$, and its fractal dimension is $\frac{\log 6}{\log 3}\sim1.631$, which is larger than $\frac{\log3}{\log2}\sim 1.585$. Then, one may naturally expect that this generalization gives a fractal code with $k\sim O(L)$ and $d \sim O(L^{\frac{\log 6}{\log 3}})$ where $k$ is the number of encodable three-dimensional logical spins. 

The key observation here is that the fractal dimension of Sierpinski triangle grows as the inner dimension of spins increases. In particular, at the limit where $p$ goes to infinity, we notice 
\begin{align}
\mathcal{D}^{(2)}_{p} = \frac{\log(\frac{p(p+1)}{2})}{\log p} \rightarrow 2 \qquad \mbox{for} \quad p\rightarrow \infty.
\end{align}
Therefore, by taking sufficiently large $p$, one can construct a fractal code with $k^{(p)} \sim O(L)$ and $d \geq O(L^{2-\epsilon})$ for an arbitrary small $\epsilon>0$ where $k^{(p)}$ is the number of encodable $p$-dimensional spins. This family of fractal codes based on generalized Sierpinski triangle will saturate the bound in Eq.~(\ref{eq:bound}) asymptotically. While our construction of fractal codes uses $p$-dimensional spins with $p>2$, one can simulate these fractal codes through two-dimensional spins. 

Then, what about the bound on higher-dimensional systems with $D>2$ ? Fortunately, there exist higher-dimensional generalizations of Sierpinski triangle constructed on a $D$-dimensional hypercubic lattice (see~\cite{Wolfram_Text} for example). For $D$-dimensional Sierpinski triangle with $p$-dimensional spins, its fractal dimension is given by
\begin{align}
\mathcal{D}^{(D)}_{p}= \log \left( \frac{p(p+1)\cdots(p+D-1)}{D!} \right)/\log(p)
\end{align}
which approaches to $D$ as $p$ goes to infinity: $\mathcal{D}^{(D)}_{p}\rightarrow D$ for $p \rightarrow \infty$. A fractal code based on $D$-dimensional Sierpinski triangle has $k^{(p)} \sim O(L^{D-1})$ and $d \sim O(L^{\mathcal{D}^{(D)}_{p}})$, and one can construct fractal codes which saturate the bound asymptotically in any spatial dimension. \\

\tb{Main theorem:}
Discussion above is valid only if the assumption that the fractal dimension of the code distance is equal to the fractal dimension of Sierpinski triangle is true:

\begin{theorem}[Fractal dimension of code distance]\label{theorem_fractal}
In fractal codes, the fractal dimension of the code distance $d$ is equal to the fractal dimension of the Sierpinski triangle:
\begin{align}
k \sim O(L^{D-1}) \qquad d \sim O(L^{\mathcal{D}^{(D)}_{p}})
\end{align}
where $\mathcal{D}^{(D)}_{p}$ is the fractal dimension of $D$-dimensional Sierpinski triangle constructed with $p$-dimensional spins, and $k$ is the number of encodable logical $p$-dimensional spins.
\end{theorem}

The rest of the paper is dedicated to the proof of this theorem. A precise definition of fractal codes is presented in subsequent sections, and a definition of local codes is presented in appendix~\ref{sec:intro} along with a brief introduction to theory of error-correcting codes and a derivation of the local code bound.

In deriving the local code bound, only the locality of interaction terms on a discretized space is assumed. It is interesting to observe that, from such a simple assumption, an area law naturally arises in fractal codes; the number of encoded bits $k$ is area-like with $k\sim O(L^{D-1})$, while the code distance $d$ is asymptotically volume-like with $d \sim O(L^{D-\epsilon})$. It is also interesting to note that, the area-law arising in fractal codes can be derived from purely classical calculations while derivations of black hole area-law and entanglement entropy area-law both require quantum mechanics. Indeed, fractal codes can be represented as a spin network state~\cite{Penrose71}. In a very broad and expanded sense, fractal codes, and other physical realizations of cellular automaton, are black-hole like since their inner states are completely determined by the degree of freedom at the surface. However, a connection between fractal codes and black holes has not been established, with further work needed. \\

\tb{Comments:}
The paper is organized as follows. In section~\ref{sec:2dim}, we discuss two-dimensional cases. In section~\ref{sec:3dim}, we discuss three-dimensional cases. In section~\ref{sec:Ddim}, we sketch the proof for $D>3$. In section~\ref{sec:open}, we list possible future problems and give some discussion. The paper is written in a self-consistent way, and most of non-trivial mathematical proofs are presented in appendix~\ref{sec:proof}, so we hope that the main discussion is accessible to readers both in coding theory and physics community. The main technical result of this paper is lemma~\ref{lemma_ineq} concerning the weights of raw vectors of the Sierpinski triangle, which may be of interest by its own. 

\section{Two-dimensional fractal code}\label{sec:2dim}

In this section, we introduce fractal codes on a two-dimensional lattice and show the asymptotic saturation of the local code bound for $D=2$. Theoretical tools to compute the code distance of fractal codes are also developed.

\subsection{Basic properties of the Sierpinski triangle}

\begin{figure}[htb!]
\centering
\includegraphics[width=0.85\linewidth]{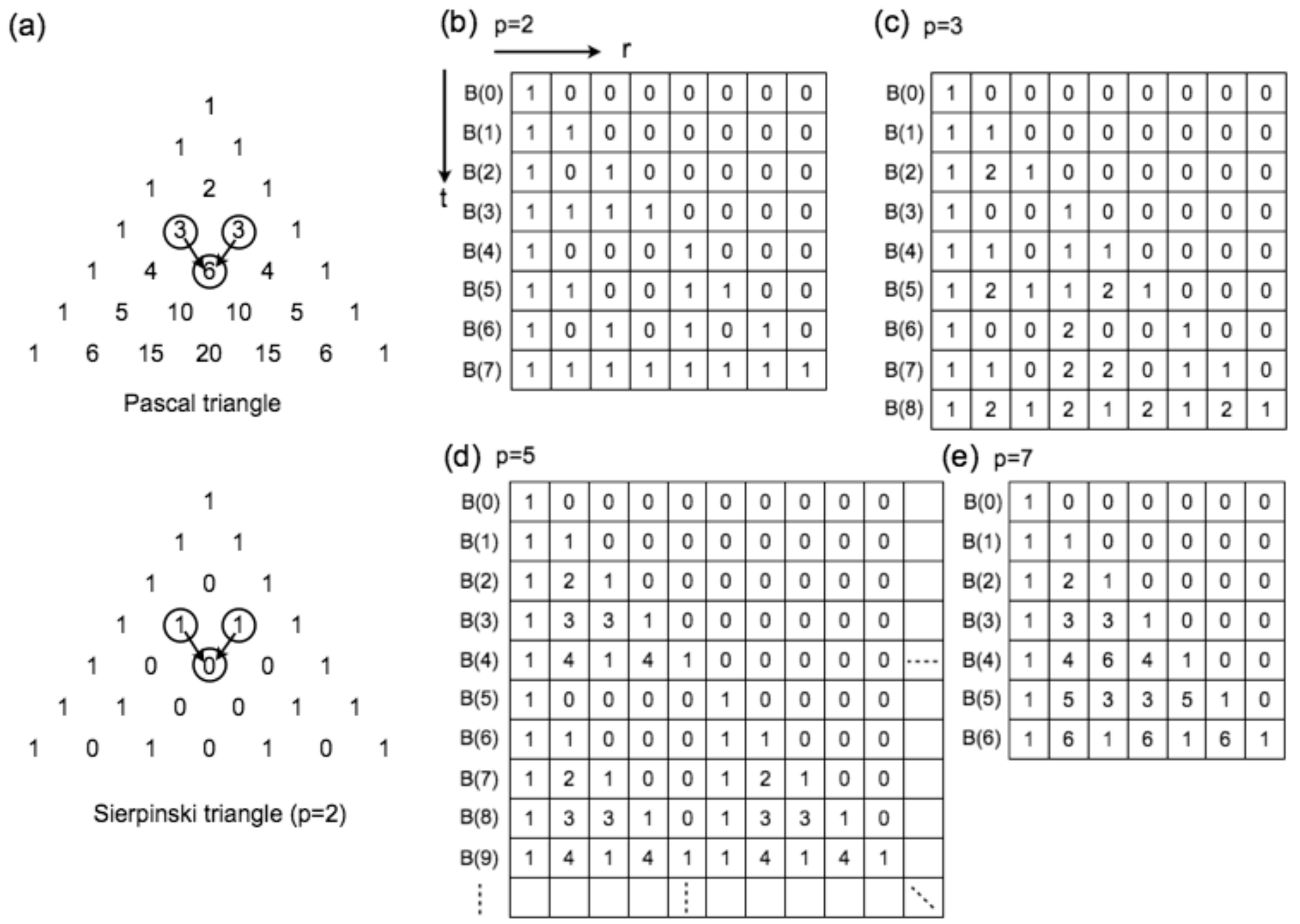}
\caption{(a) The Pascal triangle and the Sierpinski triangle. (b)-(e) The Pascal matrices. (b) $p=2$ and $m=3$. (c) $p=3$ and $m=2$. (d) $p=5$. (e) $p=7$ and $m=1$.
} 
\label{fig_entries_2D}
\end{figure}

We begin by recalling basic properties of the Sierpinski triangle~\cite{Wolfram_Text}. The Sierpinski triangle arises by considering the Pascal triangle, a triangular array of the binomial coefficients, represented modulo $p$ (see Fig.~\ref{fig_entries_2D}(a)). For our purpose, it is convenient to represent the Sierpinski triangle as a matrix (Fig.~\ref{fig_entries_2D}(b)-(e)). Consider an $L\times L$ matrix $\tb{B}$ where $L=p^{m}$ with arbitrary prime $p$ and positive integer $m$. Entries of $\tb{B}$ are denoted as $B(t)_{r}$ which corresponds to an entry at $t$-th raw and $r$-th column of the matrix for $t,r=0,\cdots,L-1$. Note that $t$ and $r$ run from $0$ to $L-1$, instead of running from $1$ to $L$ in our notation. Then, the Sierpinski triangle arises by taking the following entries:
\begin{align}
B(t)_{r}={}_tC_{r} \qquad (\mbox{mod $p$})
\end{align}
where ${}_tC_{r}=0$ for $r>t$. We call $\tb{B}$ the \emph{Pascal matrix} due to its resemblance to the Pascal triangle. 

Entries $B(t)_{r}$ obey the following constraint:
\begin{align}
B(t+1)_{r+1} = B(t)_{r} + B(t)_{r+1} \qquad \mbox{(mod $p$)} \label{eq:rule_}
\end{align}
where periodic boundary conditions are set for $r$, meaning that $B(t)_{L}=B(t)_{0}$. The entire system can be viewed as a ``computational machine'' which computes a vector $B(t)=(B(t)_{0},B(t)_{1},\cdots, B(t)_{L-1})$ at time $t$ for a given initial condition $B(0)=(1,0, \cdots, 0)$ after the ``time-evolution'' according to Eq.~(\ref{eq:rule_}). In this light, the Sierpinski triangle can be viewed as a history of time-evolution of one-dimensional cellular automaton embedded in a two-dimensional space. 
\\

\tb{Fractal dimensions:}
Fractal dimensions of the Sierpinski triangle can be computed by counting the number of non-zero entries in $\tb{B}$. For this purpose, it is useful to represent $t$ and $r$ in p-adic forms:
\begin{align*}
&r=(r_{m},r_{m-1},\cdots, r_{1})_{p}, \qquad r = \sum_{m'=1}^{m}p^{m'-1}r_{m'}\\
&t=(t_{m},t_{m-1},\cdots, t_{1})_{p}, \qquad t = \sum_{m'=1}^{m}p^{m'-1}t_{m'}
\end{align*}
where $r_{j}$ and $t_{j}$ are positive integers with $0 \leq r_{j}, t_{j} \leq p-1$. Then, entries $B(t)_{r}$ can be calculated by the following lemma:

\begin{lemma}\label{lemma_entry}
One has
\begin{align}
{}_tC_{r} = \prod_{m'=1}^{m}{}_{t_{m'}}C_{r_{m'}} \qquad (\mbox{mod $p$}),
\end{align}
and
\begin{align}
{}_tC_{r} \not= 0 \qquad (\mbox{mod $p$})\qquad \mbox{iff} \qquad
t_{m'} \geq r_{m'} \qquad \mbox{for all $m'$}.
\end{align}
\end{lemma}

The proof of the lemma is straightforward. As a direct consequence of the lemma, one can compute the fractal dimensions of the Sierpinski triangle:

\begin{corollary}[Fractal dimension]\label{corollary_entry}
Let $W(\tb{B})$ denote the number of non-zero entries in $\tb{B}$. Then, one has
\begin{align}
W(\tb{B})= \left(\frac{p(p+1)}{2}\right)^{m} = L^{\mathcal{D}_{p}^{(2)}}
\end{align}
where $L=p^{m}$, and the fractal dimension of the Pascal matrix $\tb{B}$ is given by
\begin{align}
\mathcal{D}_{p}^{(2)}= \log \left(\frac{p(p+1)}{2} \right)/\log p.
\end{align}
\end{corollary}

\begin{figure}[htb!]
\centering
\includegraphics[width=0.85\linewidth]{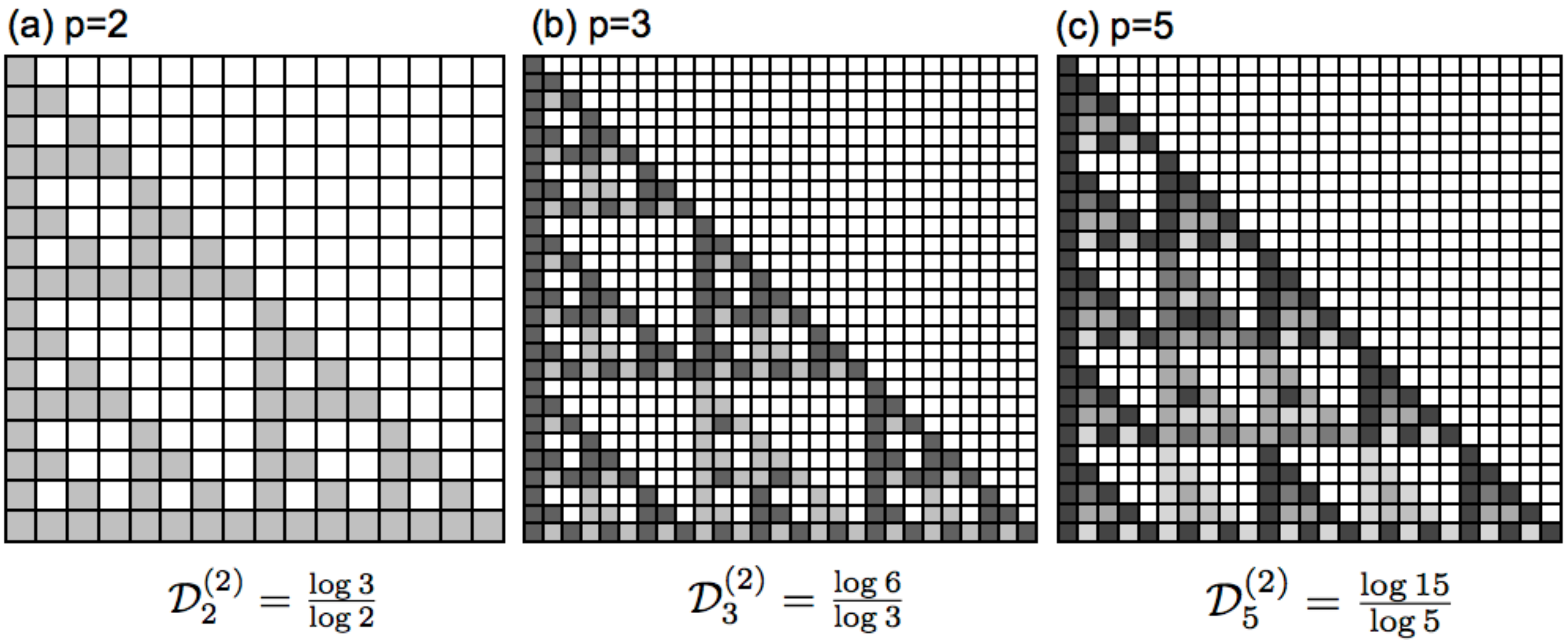}
\caption{Examples of fractal dimensions for $p=2,3,5$.
} 
\label{fig_fractal}
\end{figure}

\begin{proof}
The number of non-zero entries in $\tb{B}$ is equal to the number of pairs of $t$ and $r$ such that
\begin{align*}
t_{m'} \geq r_{m'} \qquad \mbox{for all $m'$}
\end{align*}
from lemma~\ref{lemma_entry}. There are $\frac{p(p+1)}{2}$ possible pairs of $(t_{m'}, r_{m'})$ satisfying $t_{m'} \geq r_{m'}$ for each $m'$. Therefore, in total, there are $W(\tb{B})=\left(\frac{p(p+1)}{2}\right)^{m}$ non-zero entries. 
\end{proof}

Some examples of the Sierpinski triangle and its fractal dimensions are shown in Fig.~\ref{fig_fractal}.\\

\begin{figure}[htb!]
\centering
\includegraphics[width=0.5\linewidth]{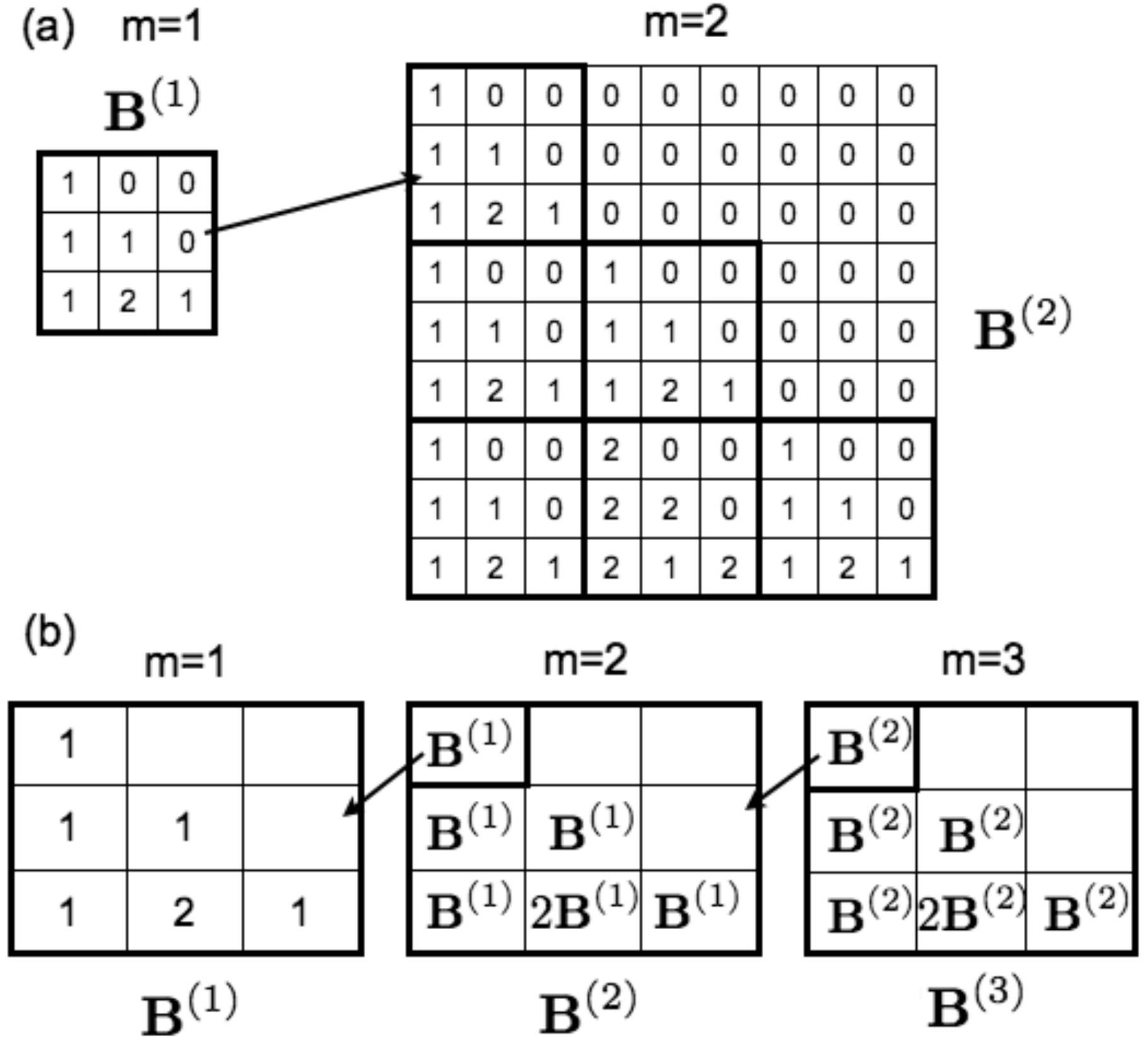}
\caption{(a) An example of a self-similar property for $p=3$. $\tb{B}^{(1)}$ appears repeatedly as submatrices of $\tb{B}^{(2)}$. (b) Self-similar properties at different length scales. 
} 
\label{fig_self_similar}
\end{figure}

\tb{Self-similarity:}
The Pascal matrices $\tb{B}$ have fractal properties with \emph{self-similar structures}. In particular, as shown in Fig.~\ref{fig_self_similar}(a), similar patterns appear repeatedly at various length scales. Self-similarity of the Sierpinski triangle is summarized as follows:

\begin{fact}[Self-similarity]\label{fact_self-similar}
We denote the Pascal matrix defined for $L=p^{m}$ as $\tb{B}^{(m)}$. Then, one has
\begin{align}
\tb{B}^{(1)}= \begin{bmatrix}
{}_0C_{0} ,&  {}_0C_{1} ,& \cdots ,&{}_0C_{p-1}\\
{}_1C_{0} ,&  {}_1C_{1} ,& \cdots ,&{}_1C_{p-1}\\
\vdots     &   \vdots    & \vdots & \ddots \\
{}_{p-1}C_{0} ,&  {}_{p-1}C_{1} ,& \cdots ,&{}_{p-1}C_{p-1}
\end{bmatrix}
\end{align}
and
\begin{align}
\tb{B}^{(m)}= \begin{bmatrix}
{}_0C_{0}\cdot \tb{B}^{(m-1)} ,& {}_0C_{1}\cdot \tb{B}^{(m-1)},& \cdots ,& {}_0C_{p-1}\cdot \tb{B}^{(m-1)} \\
{}_1C_{0}\cdot \tb{B}^{(m-1)} ,& {}_1C_{1}\cdot \tb{B}^{(m-1)},& \cdots ,& {}_1C_{p-1}\cdot \tb{B}^{(m-1)} \\
\vdots & \vdots & \vdots & \ddots \\
{}_{p-1}C_{0}\cdot \tb{B}^{(m-1)} ,& {}_{p-1}C_{1}\cdot \tb{B}^{(m-1)},& \cdots ,& {}_{p-1}C_{p-1}\cdot \tb{B}^{(m-1)} \\
\end{bmatrix}.
\end{align}
\end{fact}

Therefore, small Pascal matrices $\tb{B}^{(m-1)}$ appear repeatedly as submatrices of the original Pascal matrix $\tb{B}^{(m)}$. Fact~\ref{fact_self-similar} can be proven easily by lemma~\ref{lemma_entry}. It is worth looking at an example for $p=2$:
\begin{align}
\tb{B}^{(1)}= \begin{bmatrix}
1 ,& 0 \\
1 ,& 1
\end{bmatrix},\qquad
\tb{B}^{(m)}= \begin{bmatrix}
\tb{B}^{(m-1)} ,& \tb{0} \\
\tb{B}^{(m-1)} ,& \tb{B}^{(m-1)}
\end{bmatrix}
\end{align}
where $\tb{0}$ represents a $2^{m-1}\times 2^{m-1}$ zero matrix. An example for $p=3$ is shown in Fig.~\ref{fig_self_similar}. 

\subsection{Definition of fractal codes}

\begin{figure}[htb!]
\centering
\includegraphics[width=0.7\linewidth]{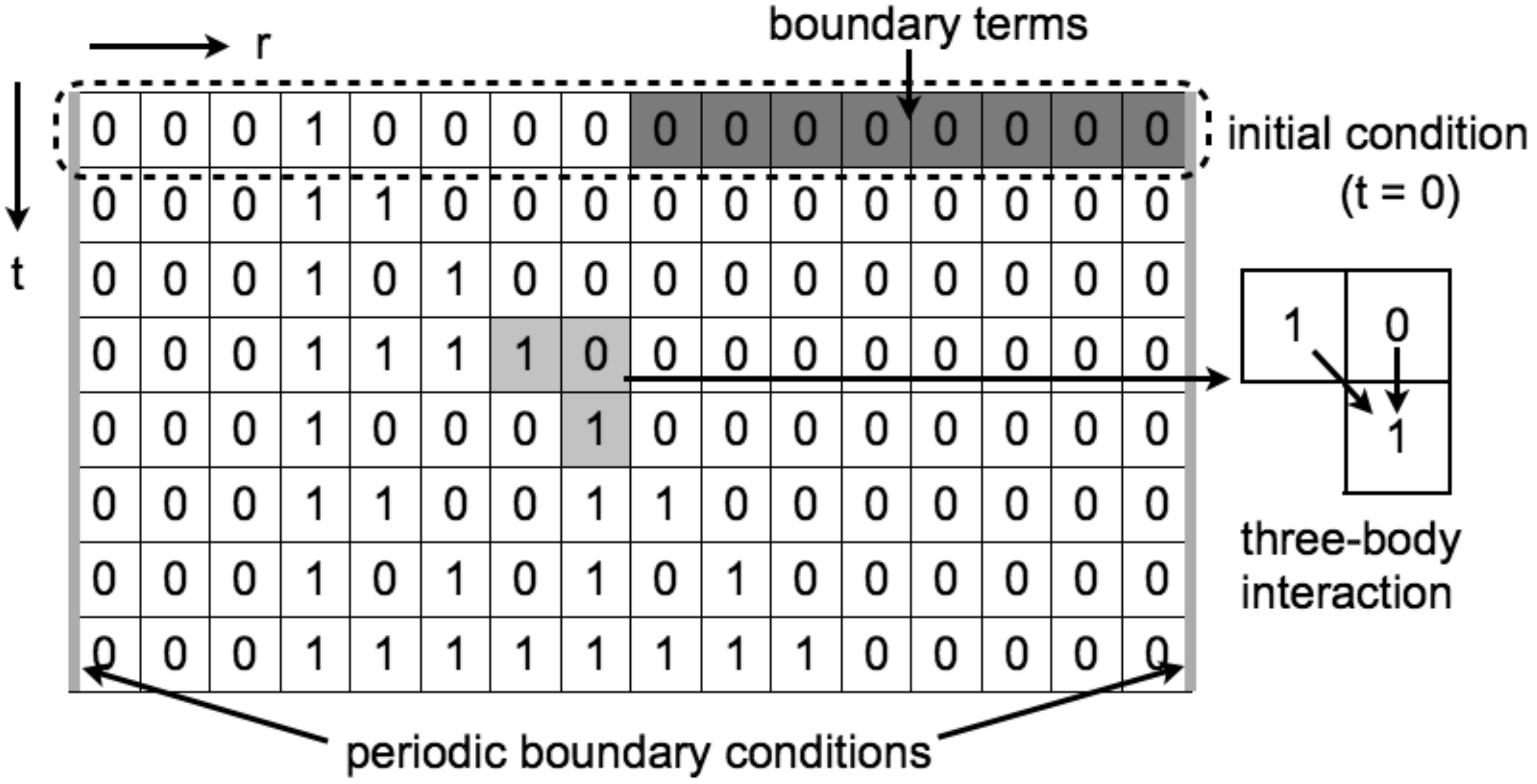}
\caption{The construction of fractal codes. The example above shows the case with $p=2$ and $L=8$ ($m=3$). Periodic boundary conditions are set along the time axis. Admissible spin configurations of a fractal code appear as ground states of a three-body Hamiltonian. The first raw at $t=0$ is called an initial condition. Eight spins on the right hand side of the initial condition are zero due to boundary terms. 
} 
\label{fig_CA}
\end{figure}

Next, we give a precise definition of fractal codes in two-dimensional systems. Consider a two-dimensional square lattice with $n = L\times 2L$ spins where spins are $p$-dimensional and spin values are $0,\cdots,p-1$. We assume that $p$ is a \emph{prime number}, and $L=p^{m}$ with arbitrary positive integer $m$. Each spin is labeled by ``time'' $t$ and ``position'' $r$ where $t=0,\cdots, L-1$ and $r=0,\cdots, 2L-1$. We set periodic boundary conditions along the time axis, and set open boundary conditions along the position axis  (see Fig.~\ref{fig_CA}).

The admissible spin configurations of fractal codes obeys the following local constraint:
\begin{align}
x(t+1)_{r} = x(t)_{r-1} + x(t)_{r} \qquad (\mbox{mod $p$})\qquad 0 \leq t \leq L-2 \label{eq:rule}
\end{align}
where $x(t)_{r}=0,\cdots,p-1$ represents the spin value at $(t,r)$. Notice that such spin configurations can be physically realized as ground states of the following three-body local Hamiltonian:
\begin{align}
H_{fractal} = \sum_{t,r} \Pi(t)_{r}, \qquad \Pi(t)_{r} = x(t+1)_{r} - x(t)_{r-1} - x(t)_{r} \qquad (\mbox{mod $p$}) 
\end{align}
with a finite energy gap. There are $p^{2L}$ admissible spin configurations which can be uniquely specified by the ``initial condition'' $x(0)=( x(0)_{0},\cdots,x(0)_{2L-1} )$ for $t=0$ on the first raw of a lattice (see Fig.~\ref{fig_CA}). The original Sierpinski triangle arises by taking  $x(0)=(1,0, \cdots,0 )$.

Now, we construct the fractal codes based on admissible spin configurations obeying Eq.~(\ref{eq:rule}). Here, we further limit our considerations to spin configurations which satisfy the following initial condition:
\begin{align}
x(0)_{r}=0 \qquad \mbox{for}\quad r \geq L.\label{eq:boundary}
\end{align}
This constraint may be physically realized by setting additional terms on the boundary of the lattice:
\begin{align}
H_{boundary} = \sum_{r \geq L} x(0)_{r}.
\end{align}
We denote a space of spin configurations specified by Eq.~(\ref{eq:rule}) and Eq.~(\ref{eq:boundary}) as $\mathcal{C}^{(2)}_{p}$, and call it the \emph{codeword space} of a fractal code. Coding properties of fractal codes are summarized in the following theorem. 

\begin{theorem}[Two-dimensional fractal code]\label{theorem_main_fractal_1}
For the codeword space $\mathcal{C}^{(2)}_{p}$ specified by Eq.~(\ref{eq:rule}) and Eq.~(\ref{eq:boundary}), let $k$ be the number of encodable $p$-dimensional spins and $d$ be the code distance of the code (i.e. the minimal Hamming distance among all the possible spin configurations). Then, we have
\begin{align}
k=L,\qquad
d = L^{\mathcal{D}^{(2)}_{p}}.
\end{align}
where
\begin{align}
\mathcal{D}^{(2)}_{p}= \log \left(\frac{p(p+1)}{2} \right)/\log(p).
\end{align}
\end{theorem}

Here, we notice that $\mathcal{D}^{(2)}_{p}$ increases as $p$ increases. In particular, since $\mathcal{D}^{(2)}_{p} \rightarrow 2$ for $p \rightarrow \infty$, we can construct a code which asymptotically saturates the bound $k \sqrt{d}\leq O(n)$ in Eq.~(\ref{eq:bound}). 

\subsection{Principal vectors}

Finally, we give the proof of theorem~\ref{theorem_main_fractal_1} by developing a theoretical tool which is useful in computing the code distance of fractal codes.\\

\tb{Principal vectors:} Let us consider the Pascal matrix $\tb{B}$. We denote entries of the $t$-th row in $\tb{B}$ as $B(t)$ where
\begin{align*}
B(t)=(B(t)_{0},\cdots,B(t)_{L-1})
\end{align*}
and call them \emph{principal vectors}. For example, with $m=2$ and $p=2$, we have the following principal vectors:
\begin{align*}
B(0)&=(1,0,0,0)\\
B(1)&=(1,1,0,0)\\ 
B(2)&=(1,0,1,0)\\ 
B(3)&=(1,1,1,1).
\end{align*}
See examples in Fig.~\ref{fig_entries_2D}. 

Note that principal vectors $B(t)$ are all independent. In particular, for an arbitrary vector $v=(v_{0},v_{1},\cdots,v_{L-1})$ with $v_{j}=0,\cdots,p-1$, one can decompose $v$ uniquely by using principal vectors:
\begin{align}
v = \sum_{t=0}^{L-1}c(t)B(t)\qquad \mbox{(mod $p$)}
\end{align}
where $c(t)=0,\cdots,p-1$. The following lemma is particularly useful in lower bounding the weight of $v$:

\begin{lemma}[Inequality on principal vectors]\label{lemma_ineq}
Consider the following linear combination of principal vectors:
\begin{align*}
v= \sum_{t=0}^{L-1}c(t)B(t)
\end{align*}
and denote the smallest positive integer $t$ such that $c(t)\not=0$ as $t_{min}$. Then, one has
\begin{align}
W(v) \geq W(B(t_{min}))
\end{align}
where $W(v)$ represents the number of non-zero entries in $v$.
\end{lemma}

Below, we give an intuition on the proof for $p=2$ with an example. Consider the case with $p=2$ and $L=8$. One may easily see that the lemma holds for the following vectors:
\begin{align*}
&B(2)=(1,0,1,0,0,0,0,0), \qquad B(5)=(1,1,0,0,1,1,0,0)\\
&B(2) + B(5) = (0,1,1,0,1,1,0,0)
\end{align*}
since $W(B(2)+B(5)) \geq W(B(2))$. An important observation is that the first four entries and the last four entries of $B(5)$ are exactly the same; $(1,1,0,0)(1,1,0,0)$, while $B(2)$ have non-zero entries only on the first four; $(1,0,1,0)(0,0,0,0)$. Then, even if some entries of $B(2)$ were cancelled by adding $B(5)$ on the first four entries, these eliminated entries would be recovered on the last four entries as a result of adding $B(5)$. By generalizing this observation, one notices that adding $B(t)$ ($t\geq4$) to $v$ does not decrease the weight of $v$ if the last four entries of $v$ are all zero, since $B(t)$ has the same entries for the first and last four entries. Note that such $v$ can be written as a linear combination of $B(0),B(1),B(2),B(3)$. In fact, one can show that adding $B(t)$ to $v$ such that $t>t_{min}$ never decreases the weight: $W(v+B(t)) \geq W(v)$ by a similar reasoning along with self-similar properties of the Sierpinski triangle. Therefore, one obtains $W(v) \geq W(B(t_{min}))$ for $p=2$. 

It is worth looking at another example for $p=3$ and $m=2$:
\begin{align*}
&B(2) =(1,2,1,0,0,0,0,0,0),\quad B(5)=(1,2,1,1,2,1,0,0,0),\quad B(7)=(1,1,0,2,2,0,1,1,0)\\
&B(2)+B(5)+B(7)= (0,2,2,0,1,1,1,1,0).
\end{align*}
Then, we notice 
\begin{align*}
W(B(2)+B(5)+B(7)) \geq W(B(2))
\end{align*}
and the lemma holds. The proof for $p>2$ is non-trivial, and is presented in appendix~\ref{appendix:A2}. 
\\

\tb{Principal vectors for fractal codes:}
Note that the Sierpinski triangle $\tb{B}$ and principal vectors $B(t)$ appear when one chooses the following initial condition in fractal codes:
\begin{align*}
x(0)=(1,0,0, \cdots)
\end{align*}
where the entire spin configuration may be represented as an $L \times 2L$ matrix:
\begin{align*}
\begin{bmatrix}
\tb{B} , \tb{0} 
\end{bmatrix} = 
\begin{bmatrix}
B(0), & \vec{0} \\
B(1), & \vec{0} \\
\vdots  &\vdots \\
B(L-1),& \vec{0}
\end{bmatrix}
\end{align*}
where $\tb{0}$ represents an $L\times L$ zero matrix and $\vec{0}$ represents an $L$-component zero vector.

While the fractal codes are defined as an $L\times 2L$ matrix for $t=0,\cdots, L-1$, one may naturally generalize the definition of fractal codes for $t=0,\cdots, 2L-1$ as a $2L\times 2L$ matrix. Then, the spin configuration generated from $x(0)=(1,0,0, \cdots)$ can be expressed as the following $2L\times 2L$ matrix:
\begin{align}
\begin{bmatrix}
\tb{B} ,& \tb{0} \\
\tb{B} ,& \tb{B}
\end{bmatrix} = 
\begin{bmatrix}
B(0), & \vec{0} \\
\vdots  &\vdots \\
B(L-1),& \vec{0} \\
B(0), & B(0) \\
\vdots  &\vdots \\
 B(L-1), & B(L-1)
\end{bmatrix}
\end{align}
where the $(t+L)$-th raws are given by $(B(t),B(t))$ due to Fact.~\ref{fact_self-similar}. For a later purpose, it is convenient to redefine principal vectors as $2L$-component vectors instead of $L$-component vectors:
\begin{equation}
\begin{split}
B(t) &\leftarrow (B(t),  \vec{0}) \\
B(t+L) &\leftarrow  (B(t), B(t))
\end{split}
\end{equation}
for $t=0,\cdots,L-1$, obtaining a complete set of $2L$ principal vectors. Note that these redefined principal vectors are all independent, and still obey lemma~\ref{lemma_ineq}. \\

\tb{Time evolution:}
We have analyzed a spin configuration arising from an initial condition $x(0)=(1,0,0,0,\cdots)$. Redefined $2L$-component principal vectors can be used for decomposing arbitrary initial conditions in fractal codes:
\begin{align}
x(0)=\sum_{t=0}^{2L-1} c(t)B(t)\qquad (\mbox{mod $p$}).
\end{align}
Recall that $x(0)_{r}=0$ for $r \geq L$ due to the boundary condition in Eq.~(\ref{eq:boundary}), and thus, $c(t)=0$ for $t\geq L$. Then, the initial condition can be decomposed as follows
\begin{align}
x(0)=\sum_{t=0}^{L-1} c(t)B(t)\qquad (\mbox{mod $p$})
\end{align}
by using $B(t)$ with $t=0,\cdots,L-1$ only. Therefore, the $t$-th raw $x(t)$ can be represented as follows
\begin{align}
x(t)=\sum_{\tau=0}^{L-1} c(\tau)B(\tau+t)\qquad (\mbox{mod $p$})
\end{align}
since the time evolution rule of fractal codes is linear.\\

\tb{Code distances:}
Finally, we prove theorem~\ref{theorem_main_fractal_1}. In order to show $d=L^{\mathcal{D}^{(2)}_{p}}$, one needs to prove that the minimal Hamming distance between all the pairs of codewords is equal to $L^{\mathcal{D}^{(2)}_{p}}$. This problem can be simplified further since fractal codes are \emph{linear}. Let us represent the spin configuration generated from an initial condition $v$ as $\tb{V}(v)$. Then, the Hamming weight between two spin configurations $\tb{V}(v)$ and $\tb{V}(v')$ is given by
\begin{align}
W(\tb{V}(v) + \tb{V}(v')) = W(\tb{V}(v+v')) 
\end{align}
where $v+v'$ is computed modulo $p$. Since fractal codes are linear:
\begin{align}
\tb{V}(v), \tb{V}(v') \in \mathcal{C}_{p}^{(2)} \ \rightarrow \ \tb{V}(v+v')\in \mathcal{C}_{p}^{(2)}
\end{align}
where $\mathcal{C}_{p}^{(2)}$ is the codeword space, one only needs to prove 
\begin{align}
d \equiv \min_{v\not=\vec{0}} W(\tb{V}(v)) = L^{\mathcal{D}^{(2)}_{p}}
\end{align}
by finding a spin configuration $\tb{V}(v)$ with the lowest weight.

The weight of a spin configuration $V(x(0))$ generated from an initial condition $x(0)$ is given by
\begin{align*}
W(\tb{V}(x(0))) = \sum_{t=0}^{L-1} W(x(t)).
\end{align*}
We denote the smallest $t$ such that $c(t)\not=0$ as $t_{min}$. Then, due to lemma~\ref{lemma_ineq}, we have
\begin{align*}
W(x(t)) \geq W(B(t_{min}+t)),
\end{align*}
which leads to
\begin{align*}
\sum_{t=0}^{L-1} W(x(t)) \geq \sum_{t=0}^{L-1} (B(t_{min}+t)).
\end{align*}
Since $W(B(t+L))=2W(B(t))$ for $t \geq L$ due to the self-similarity, we have 
\begin{align*}
\sum_{t=0}^{L-1} (B(t_{min}+t)) \geq \sum_{t=0}^{L-1} (B(t)) = L^{\mathcal{D}_{p}^{(2)}}.
\end{align*}
The bound is tight for $x(0)=(1,0,\cdots)$. This completes the proof of theorem~\ref{theorem_main_fractal_1}.\\

\tb{Comments:}
The reason why we limit our considerations to spin configurations obeying the boundary term Eq.~(\ref{eq:boundary}) comes from a certain technical difficulty. For $p=2$ and an initial condition $( x(0)_{0},\cdots,x(0)_{2L-1} ) =(1,\cdots,1)$, the resulting spin configurations are $( x(t)_{0},\cdots,x(t)_{2L-1} ) =(0,\cdots,0)$ for $t>0$ which would lead to $d=2L$. To avoid this difficulty, we need Eq.~(\ref{eq:boundary}). This issue is closely related to the irreversibility of cellular automaton. 

\section{Three-dimensional fractal code}\label{sec:3dim}

The construction of two-dimensional fractal codes can be generalized to higher-dimensional systems ($D>2$) straightforwardly. In this section, we introduce the three-dimensional version of fractal codes and show the asymptotic saturation of the local code bound for $D=3$. 

\subsection{Basic properties of the three-dimensional Sierpinski triangle}

\begin{figure}[htb!]
\centering
\includegraphics[width=0.45\linewidth]{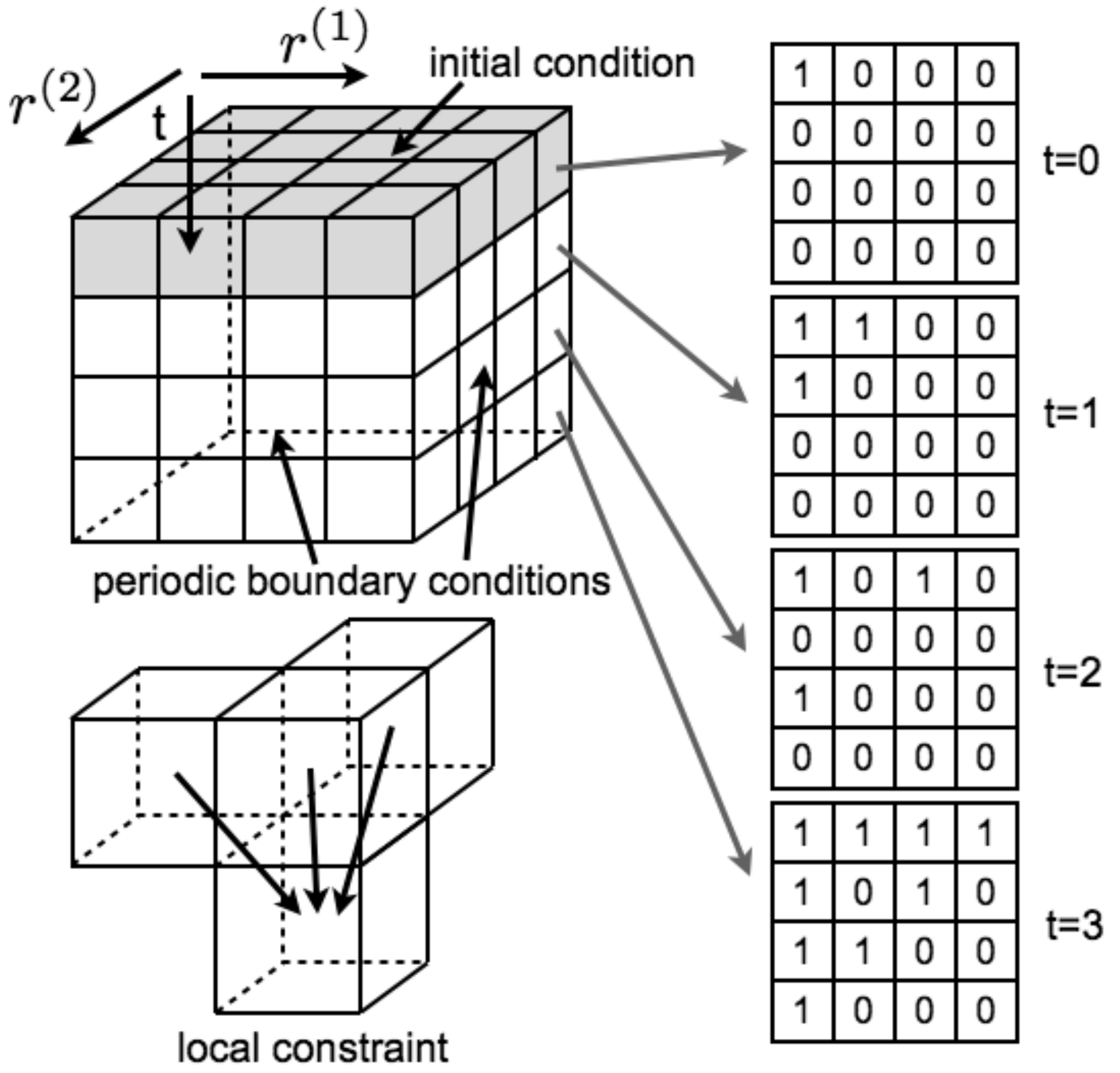}
\caption{Three-dimensional Sierpinski triangle. The example above shows the case with $p=2$ and $m=2$.
} 
\label{fig_3D_fractal}
\end{figure}

We begin by recalling basic properties of the three-dimensional Sierpinski triangle. It is convenient to represent the Sierpinski triangle as an $L\times L \times L$ tensor with $L=p^{m}$, denoted by $\tb{B}$ (see Fig.~\ref{fig_3D_fractal}). Entries of $\tb{B}$ are denoted as $B(t)_{r^{(1)},r^{(2)}}$ for $t,r^{(1)},r^{(2)}=0,\cdots,L-1$. Then, the Sierpinski triangle arises by taking the following entries:
\begin{equation}
\begin{split}
B(t)_{r^{(1)},r^{(2)}} = {}_{t} C _{r^{(1)}} \cdot {}_{t-r^{(1)}} C_{r^{(2)}} = \frac{t !}{r^{(1)}!r^{(2)}!(t-r^{(1)}-r^{(2)})!}
\qquad \mbox{(mod $p$)}.
\end{split}
\end{equation}
Note that entries $B(t)_{r^{(1)},r^{(2)}}$ obey the following constraint:
\begin{align}
B(t+1)_{r^{(1)},r^{(2)}} = B(t)_{r^{(1)},r^{(2)}} +B(t)_{r^{(1)}-1,r^{(2)}} + B(t)_{r^{(1)},r^{(2)}-1} \qquad \mbox{(mod $p$)} \label{eq:update__}
\end{align}
where we set periodic boundary conditions for $r^{(1)}$ and $r^{(2)}$. We call $\tb{B}$ the Pascal tensor.

By denoting the $t$-th layer of $\tb{B}$ as $B(0,t)$, the Pascal tensor $\tb{B}$ can be represented as follows:
\begin{align}
\tb{B} =
\begin{bmatrix}
B(0,0) \\
B(0,1) \\
\vdots \\
B(0,L-1) 
\end{bmatrix}.
\end{align}
For example, with $p=2$ and $m=2$, one has
\begin{align*}
\tb{B} =
\begin{bmatrix}
B(0,0) \\
B(0,1) \\
B(0,2) \\
B(0,3) 
\end{bmatrix}
\end{align*}
where
\begin{align*}
B(0,0)=\begin{bmatrix}
1,& 0 ,& 0 ,& 0 \\
0,& 0 ,& 0 ,& 0 \\
0,& 0 ,& 0 ,& 0 \\
0,& 0 ,& 0 ,& 0 \\
\end{bmatrix}\  \
B(0,1)=\begin{bmatrix}
1,& 1 ,& 0 ,& 0 \\
1,& 0 ,& 0 ,& 0 \\
0,& 0 ,& 0 ,& 0 \\
0,& 0 ,& 0 ,& 0 \\
\end{bmatrix}\  \\
B(0,2)=\begin{bmatrix}
1,& 0 ,& 1 ,& 0 \\
0,& 0 ,& 0 ,& 0 \\
1,& 0 ,& 0 ,& 0 \\
0,& 0 ,& 0 ,& 0 \\
\end{bmatrix}\ \
B(0,3)=\begin{bmatrix}
1,& 1 ,& 1 ,& 1 \\
1,& 0 ,& 1 ,& 0 \\
1,& 1 ,& 0 ,& 0 \\
1,& 0 ,& 0 ,& 0 \\
\end{bmatrix}.
\end{align*}
One can view $\tb{B}$ as a history of time-evolution of an initial condition
\begin{align*}
B(0,0) = \begin{bmatrix}
1,& 0 ,& \cdots ,& 0 \\
0,& 0 ,& \cdots ,& 0 \\
\vdots & \vdots  & \ddots & 0 \\
0,& 0 ,& 0 ,& 0 \\
\end{bmatrix}
\end{align*}
according to the update rule in Eq.~(\ref{eq:update__}):
\begin{align}
B(0,0)\rightarrow B(0,1) \rightarrow \cdots \rightarrow B(0,L-1).
\end{align}

\tb{Fractal dimensions:}
To compute fractal dimensions of the Sierpinski triangle, we represent $t$, $r^{(1)}$ and $r^{(2)}$ in p-adic forms:
\begin{align*}
&r^{(1)}=(r^{(1)}_{m}r^{(1)}_{m-1}\cdots r^{(1)}_{1})_{p}, \qquad r = \sum_{m'=1}p^{m'-1}r_{m'}\\
&r^{(2)}=(r^{(2)}_{m}r^{(2)}_{m-1}\cdots r^{(2)}_{1})_{p}, \qquad r = \sum_{m'=1}p^{m'-1}r_{m'}\\
&t=(t_{m}t_{m-1}\cdots t_{1})_{p}, \quad \qquad \qquad t = \sum_{m'=1}p^{m'-1}t_{m'}.
\end{align*}
From lemma~\ref{lemma_entry}, entries $B(t)_{r^{(1)},r^{(2)}}$ can be expressed as follows:
\begin{align}
B(t)_{r^{(1)},r^{(2)}} = \prod_{m'=1}^{m} {}_{t_{m'}} C _{r^{(1)}_{m'}}\cdot {}_{t_{m'}-r^{(1)}_{m'}} C _{r^{(2)}_{m'}}\qquad \mbox{(mod $p$)}.
\end{align}
Then, one has
\begin{align}
B(t)_{r^{(1)},r^{(2)}} \not=0 \qquad (\mbox{mod $p$})\qquad \mbox{iff} \qquad
t_{m'} \geq r^{(1)}_{m'} \quad  \mbox{and} \quad t_{m'} - r^{(1)}_{m'} \geq r^{(2)}_{m'} \quad \mbox{for all $m'$}.
\end{align}
There are only $p(p+1)(p+2)/6$ possible combinations of $(t_{m'}, r^{(1)}_{m'},r^{(2)}_{m'})$ satisfying the above condition for each $m'$. Therefore, fractal dimensions of three-dimensional Sierpinski triangle are given as follows:

\begin{corollary}[Fractal dimension]\label{corollary_entry}
Let $W(\tb{B})$ denote the number of non-zero entries in $\tb{B}$. Then, one has
\begin{align}
W(\tb{B})= \left(\frac{p(p+1)(p+2)}{6}\right)^{m} = L^{\mathcal{D}_{p}^{(3)}}
\end{align}
where the fractal dimension of the Pascal matrix $\tb{B}$ is given by
\begin{align}
\mathcal{D}_{p}^{(3)}= \log \left(\frac{p(p+1)(p+2)}{6} \right)/\log p.
\end{align}
\end{corollary}

\subsection{Definition of three-dimensional fractal codes}\label{sec:fractal_high_1}

Next, we give a precise definition of three-dimensional fractal codes. We consider a three-dimensional cubic lattice with $n = L\times 2L \times 2L$ spins where spins are $p$-dimensional and $L=p^{m}$. Each spin is labeled by ``time'' $t$ and two ``positions'' $r^{(1)}$ and $r^{(2)}$ with $t=0,\cdots, L-1$ and $ r^{(1)},r^{(2)}=0,\cdots, 2L-1$. We set periodic boundary conditions on all the surfaces which are parallel to the time axis. The admissible spin configurations obey the following constraint:
\begin{align}
x(t+1)_{r^{(1)},r^{(2)}} = x(t)_{r^{(1)}-1,r^{(2)}} + x(t)_{r^{(1)},r^{(2)}-1} + x(t)_{r^{(1)},r^{(2)}} \qquad (\mbox{mod $p$}) \label{eq:rule2}
\end{align}
where $x(t)_{r^{(1)},r^{(2)}}=0,\cdots,p-1$ represents the spin value at $(t,r^{(1)},r^{(2)})$. Spin configurations may be uniquely specified by ``initial conditions'' $x(0)$ with $x(0)_{r^{(1)},r^{(2)}}=0,\cdots,p-1$, which may be considered as $2L \times 2L$ matrices. 

We further limit ourselves to spin configurations which satisfy the following initial condition:
\begin{align}
x(0)_{r^{(1)},r^{(2)}}=0 \qquad \mbox{for}\quad r^{(1)} + r^{(2)} \geq L,\label{eq:boundary2}
\end{align}
and denote a space of spin configurations specified by this condition as $C^{(3)}_{p}$. Our main result is summarized in the following theorem:

\begin{theorem}[Three-dimensional fractal code]\label{theorem_main_fractal_2}
For the codeword space $C^{(3)}_{p}$, let $k$ be the number of encodable $p$-dimensional spins and $d$ be the code distance of the code. Then, we have
\begin{align}
k=\frac{L(L+1)}{2}, \qquad    d = L^{\mathcal{D}_{p}^{(3)}}
\end{align}
where 
\begin{align}
\mathcal{D}_{p}^{(3)}= \log \left(\frac{p(p+1)(p+2)}{6} \right)/\log(p).
\end{align}
\end{theorem}

When $D=3$, the fractal dimension goes to three: $\mathcal{D}_{p}^{(3)} \rightarrow 3$ for $p \rightarrow \infty$. Therefore, the code saturates the bound $k d^{1/3}\leq O(n)$ in Eq.~(\ref{eq:bound}) for $D=3$ asymptotically. 

\subsection{Principal matrix}

Finally, we give the proof of theorem~\ref{theorem_main_fractal_2}. A key idea is to generalize the notion of principal vectors and introduce \emph{principal matrices} which will be useful in decomposing spin configurations on each layer. An inequality for principal vectors in lemma~\ref{lemma_ineq} is also generalized to an inequality for principal matrices.\\

\tb{Principal vectors:} Recall that we represented the Pascal tensor $\tb{B}$ in terms of its $t$-th layers $B(0,t)$:
\begin{align*}
\tb{B} = \begin{bmatrix}
B(0,0)\\
B(0,1)\\
\vdots \\
B(0,t)
\end{bmatrix}.
\end{align*}
Matrices $B(0,t)$ are closely related to principal vectors $B(t)$. To see this point, we further expand matrices $B(0,t)$ as follows:
\begin{align}
B(0,t) =\begin{bmatrix}
   B(0,t)_{0} \\
   B(0,t)_{1} \\
   \vdots   \\
   B(0,t)_{L-1}
   \end{bmatrix}\label{eq:shorthand}
\end{align}
where $B(0,t)_{j}$ are $L$-component vectors with
\begin{align}
B(0,t)_{j}=(B(0,t)_{0,j}, B(0,t)_{1,j},\cdots , B(0,t)_{L-1,j}).
\end{align}
For example, when $p=2$ and $m=2$, we have
\begin{align*}
B(0,3)= \begin{bmatrix}
1 ,& 1 ,& 1,& 1 \\
1 ,& 0 ,& 1,& 0 \\
1 ,& 1 ,& 0,& 0 \\
1 ,& 0 ,& 0,& 0 
\end{bmatrix},
\end{align*}
and
\begin{align*}
&B(0,3)_{0} = (1,1,1,1), \quad B(0,3)_{1} = (1,0,1,0)\\ &B(0,3)_{2} = (1,1,0,0), \quad B(0,3)_{3} = (1,0,0,0). 
\end{align*}
Therefore, one may represent $B(0,3)$ as follows:
\begin{align*}
B(0,3) =\begin{bmatrix}
   B(3) \\
   B(2) \\
   B(1)   \\
   B(0)
   \end{bmatrix}
\end{align*}
where $B(0)$, $B(1)$, $B(2)$ and $B(3)$ are principal vectors. Similarly, one has
\begin{align*}
B(0,2)= \begin{bmatrix}
1 ,& 0 ,& 1,& 0 \\
0 ,& 0 ,& 0,& 0 \\
1 ,& 0 ,& 0,& 0 \\
0 ,& 0 ,& 0,& 0 
\end{bmatrix}=
\begin{bmatrix}
B(2)\\
0 \\
B(0)\\
0
\end{bmatrix}.
\end{align*}

As examples above show, matrices $B(0,t)$ can be represented in terms of principal vectors $B(t)$:

\begin{lemma}[Principal matrix]\label{lemma_3D_self}
The $t$-th layer matrix $B(0,t)$ can be represented as
\begin{align}
B(0,t) = \begin{bmatrix}
B(t)_{0}\cdot B(t) \\
B(t)_{1}\cdot B(t-1) \\
\vdots \\ 
B(t)_{L-1}\cdot B(t-L+1) 
\end{bmatrix}
\end{align}
\end{lemma}

\begin{figure}[htb!]
\centering
\includegraphics[width=0.95\linewidth]{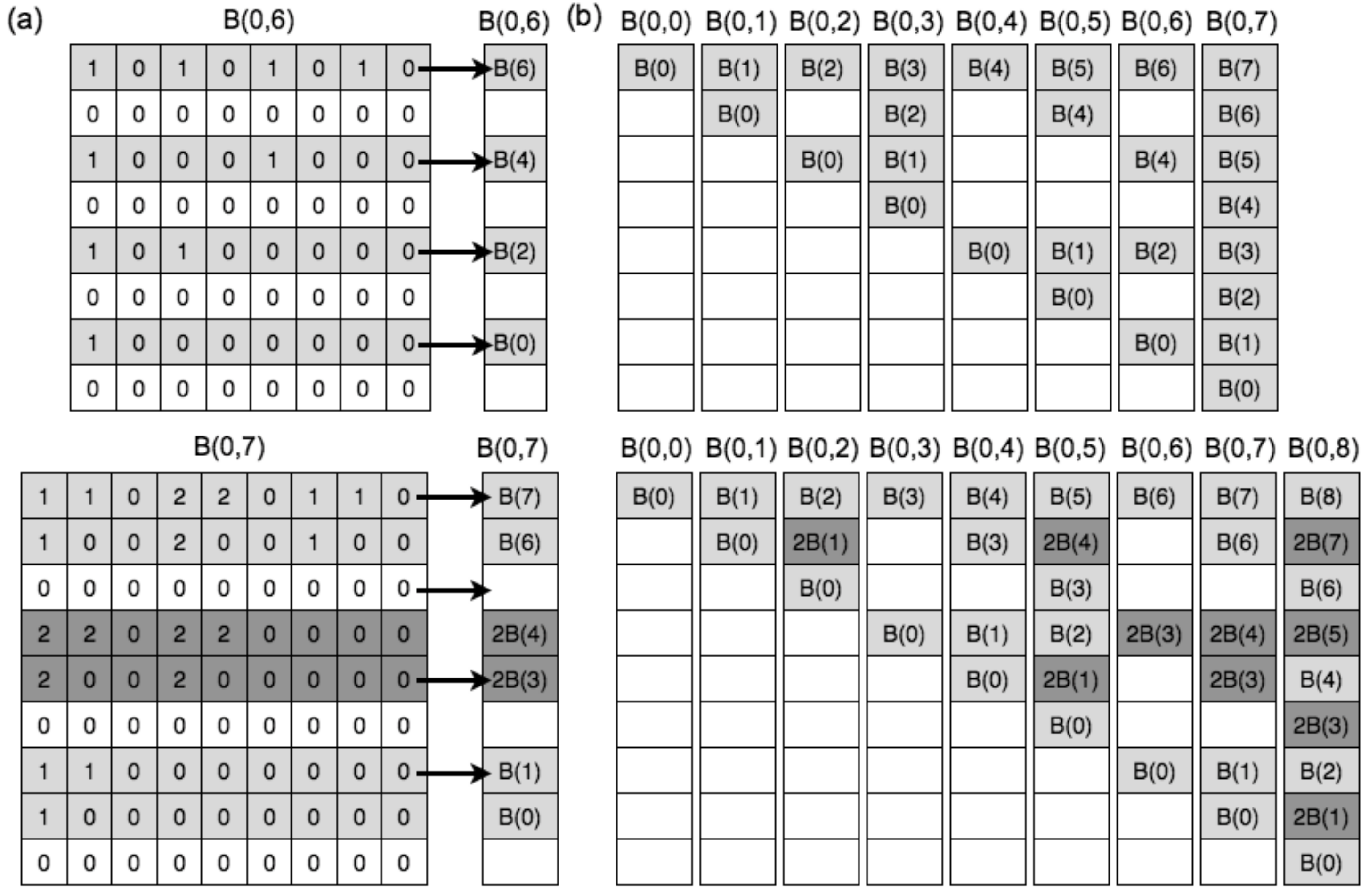}
\caption{(a) Shorthand notations of $B(0,6)$ for $p=2$ and $m=3$, and $B(0,7)$ for $p=3$ and $m=2$. (b) Principle matrices and principal vectors.
} 
\label{fig_3D_structure}
\end{figure}

As an example, let us represent $B(0,6)$ for $p=2$ and $m=3$ (see Fig.~\ref{fig_3D_structure}):
\begin{align*}
B(6)=(1,0,1,0,1,0,1,0),\qquad B(0,6)=(B(6),\vec{0},B(4),\vec{0},B(2),\vec{0},B(0),\vec{0})^{T}
\end{align*}
where $\vec{0}$ represents vectors with zero entries. Similarly, we can represent $B(0,7)$ for $p=3$ and $m=2$ as follows:
\begin{align*}
&B(7)=(1,1,0,2,2,0,1,1,0)\\ &B(0,7)=(B(7),B(6),\vec{0},2B(4),2B(3),\vec{0},B(1),B(0),\vec{0})^{T}.
\end{align*}
It is worth representing all the matrices $B(0,t)$ at once as in Fig.~\ref{fig_3D_structure}(b). In $B(0,t)$, principal vectors $B(0),\cdots, B(t)$ are distributed with weights corresponding to a principal vector $B(t)$. \\

\tb{Principal matrix:}
So far, we have analyzed the spin configuration generated by the following initial condition:
\begin{align}
B(0,0) \equiv \begin{bmatrix}
B(0) \\
0 \\
0 \\
0
\end{bmatrix}.
\end{align}
Here, we consider spin configurations generated by other initial conditions:
\begin{align}
B(a,0) \equiv \begin{bmatrix}
B(a) \\
0 \\
0 \\
0
\end{bmatrix} \qquad (\mbox{for} \ a=0,\cdots,L-1),
\end{align}
and denote the $t$-th layer of the spin configuration generated by $B(a,0)$ as $B(a,t)$. We call $B(a,t)$ \emph{principal matrices}. One may represent principal matrices $B(a,t)$ explicitly as follows:

\begin{lemma}[Principal matrix]\label{lemma_3D_self2}
A principal matrix $B(a,t)$ can be represented as
\begin{align}
B(a,t) = \begin{bmatrix}
B(t)_{0}\cdot B(t+a) \\
B(t)_{1}\cdot B(t+a-1) \\
\vdots \\ 
B(t)_{L-1}\cdot B(t+a-L+1) 
\end{bmatrix}
\end{align}
where $B(\tau + L)=2 B(\tau)$ for $0\leq \tau < L$.
\end{lemma}

Below, we show some examples. For $p=2$ and $m=2$, we have:
\begin{align*}
B(0,0) = 
\begin{bmatrix}
B(0) \\
0 \\
0 \\
0
\end{bmatrix}, \quad 
B(0,1) = 
\begin{bmatrix}
B(1) \\
B(0) \\
0 \\
0
\end{bmatrix}, \quad 
B(0,2) = 
\begin{bmatrix}
B(2) \\
0    \\
B(0) \\
0
\end{bmatrix}, \quad 
B(0,3) = 
\begin{bmatrix}
B(3) \\
B(2) \\
B(1) \\
B(0)
\end{bmatrix} \\ 
B(1,0) = 
\begin{bmatrix}
B(1) \\
0 \\
0 \\
0
\end{bmatrix}, \quad 
B(1,1) = 
\begin{bmatrix}
B(2) \\
B(1) \\
0 \\
0
\end{bmatrix}, \quad 
B(1,2) = 
\begin{bmatrix}
B(3) \\
0    \\
B(1) \\
0
\end{bmatrix}, \quad 
B(1,3) = 
\begin{bmatrix}
0    \\
B(3) \\
B(2) \\
B(1)
\end{bmatrix} \\ 
B(2,0) = 
\begin{bmatrix}
B(2) \\
0 \\
0 \\
0
\end{bmatrix}, \quad 
B(2,1) = 
\begin{bmatrix}
B(3) \\
B(2) \\
0 \\
0
\end{bmatrix}, \quad 
B(2,2) = 
\begin{bmatrix}
0    \\
0    \\
B(2) \\
0
\end{bmatrix}, \quad 
B(2,3) = 
\begin{bmatrix}
0    \\
0    \\
B(3) \\
B(2)
\end{bmatrix} \\ 
B(3,0) = 
\begin{bmatrix}
B(3) \\
0 \\
0 \\
0
\end{bmatrix}, \quad 
B(3,1) = 
\begin{bmatrix}
0    \\
B(3) \\
0 \\
0
\end{bmatrix}, \quad 
B(3,2) = 
\begin{bmatrix}
0    \\
0    \\
B(3) \\
0
\end{bmatrix}, \quad 
B(3,3) = 
\begin{bmatrix}
0    \\
0    \\
0    \\
B(3)
\end{bmatrix}.
\end{align*}

For $p=3$ and $m=1$, we have
\begin{align*}
B(0,0) = 
\begin{bmatrix}
B(0) \\
0 \\
0 
\end{bmatrix}, \quad 
B(0,1) = 
\begin{bmatrix}
B(1) \\
B(0) \\
0 
\end{bmatrix}, \quad 
B(0,2) = 
\begin{bmatrix}
B(2) \\
2B(1)    \\
B(0)
\end{bmatrix}\\ 
B(1,0) = 
\begin{bmatrix}
B(1) \\
0 \\
0 
\end{bmatrix}, \quad 
B(1,1) = 
\begin{bmatrix}
B(2) \\
B(1) \\
0 
\end{bmatrix}, \quad 
B(1,2) = 
\begin{bmatrix}
2B(0) \\
2B(2)    \\
B(1) 
\end{bmatrix}\\ 
B(2,0) = 
\begin{bmatrix}
B(2) \\
0 \\
0 
\end{bmatrix}, \quad 
B(2,1) = 
\begin{bmatrix}
2B(0) \\
B(2) \\
0 
\end{bmatrix}, \quad 
B(2,2) = 
\begin{bmatrix}
2B(1)    \\
B(0)    \\
B(2) 
\end{bmatrix}.
\end{align*}\\

\tb{Inequality for principal matrix:} One can see that principal matrices $B(a,t)$ are all independent, and an arbitrary $L\times L$ matrix can be decomposed uniquely by $B(a,t)$:
\begin{align}
v=\sum_{a,t}c(a,t)B(a,t).
\end{align} 
Here, we define the following sets:
\begin{equation}
\begin{split}
R_{0}(v)  &=\{ (a,t) : c(a,t)\not=0 \} \\
R_{1}(v)  &=\{ (a,t)\in R_{0} : a+t \leq a'+t' \ \mbox{for all} \ (a',t')\in R_{0} \} \\
R_{2}(v) &=\{ (a,t)\in R_{1} : t \leq t' \ \mbox{for all} \ (a',t')\in R_{1} \}. \\
\end{split}
\end{equation}
Note that $R_{2}\subseteq R_{1} \subseteq R_{0}$, and there is only one element in $R_{2}$. Examples of $R_{0}$, $R_{1}$ and $R_{2}$ are shown in Fig.~\ref{fig_example}. Then, for the weight of the initial condition, we have the following inequality:

\begin{figure}[htb!]
\centering
\includegraphics[width=0.45\linewidth]{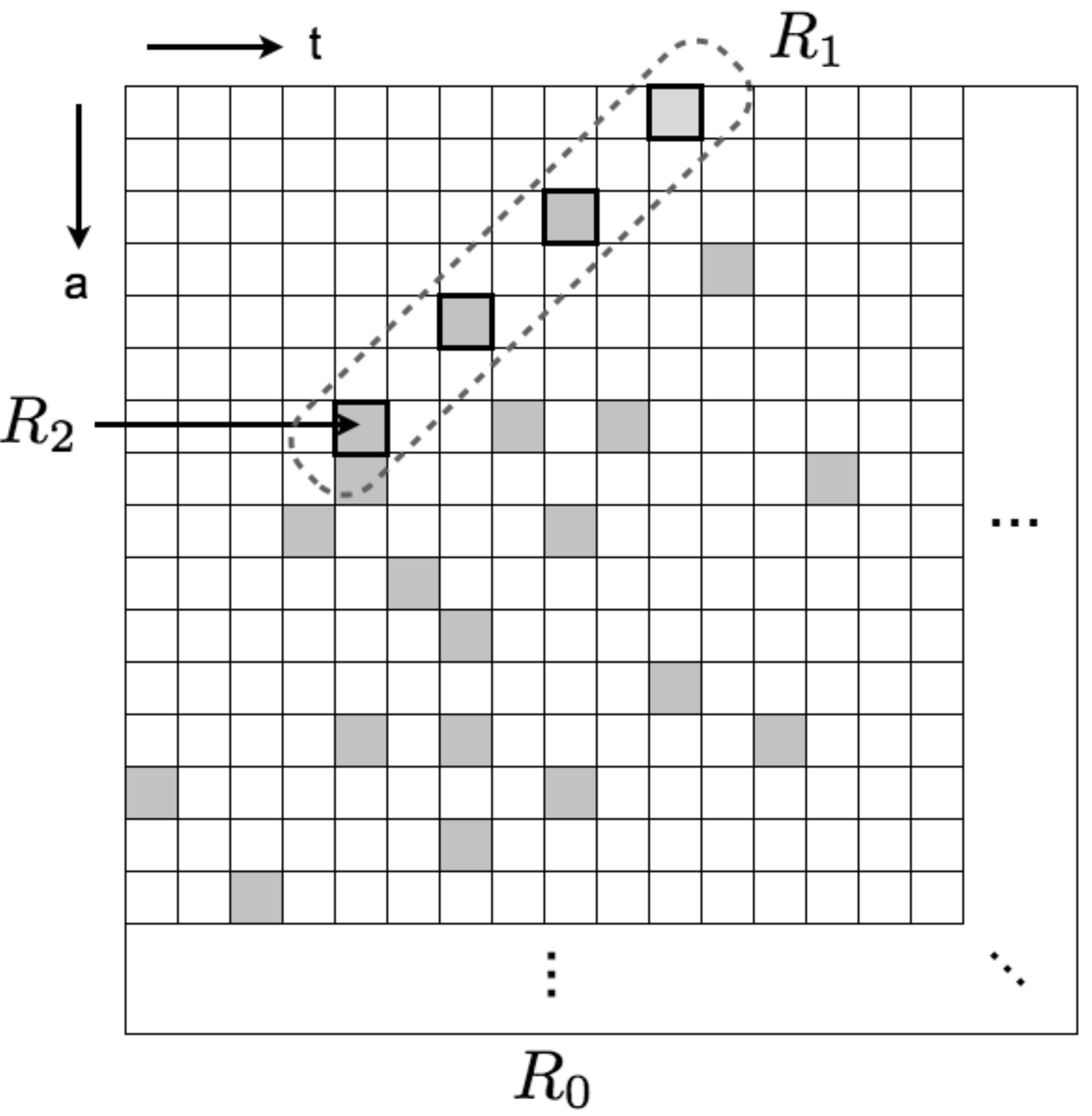}
\caption{Examples of $R_{0}$, $R_{1}$ and $R_{2}$. $R_{0}$ is a set of all shaded sites. $R_{1}$ is a set of sites with minimal $a+t$. $R_{2}$ is a subset of $R_{1}$ with minimal $t$.
} 
\label{fig_example}
\end{figure}

\begin{lemma}[Inequality on principal matrices]\label{lemma_ineq_3dim}
For a matrix
\begin{align}
v=\sum_{a,t}c(a,t)B(a,t)
\end{align} 
where $c(a,t)=0$ for all $(a,t)$ with $a+t\geq L$, let $(a',t')\in R_{2}(v)$. Then, we have
\begin{align}
W(v) \geq W\left(B(0,t')\right). 
\end{align}
\end{lemma}

As an example, let us consider the following linear decomposition:
\begin{align*}
v=B(2,3) + B(5,3) + B(1,4) + B(0,8).
\end{align*}
Then, we have
\begin{align*}
R=\{ (2,3),(5,3),(1,4),(0,8) \},\qquad R_{1} = \{ (2,3),(1,4)   \}, \qquad R_{2}=\{ (2,3) \}
\end{align*}
and
\begin{align*}
W(v) \geq W(B(0,3)). 
\end{align*}
The proof of lemma~\ref{lemma_ineq_3dim} is given in appendix~\ref{appendix:A4}.\\

\tb{Code distance:}
Finally, we prove theorem~\ref{theorem_main_fractal_2}. One can naturally extend the definition of principal matrices $B(a,t)$ for $2L\times 2L$ matrices by considering $L\times 2L \times 2L$ fractal codes. Then, lemma~\ref{lemma_ineq_3dim} holds for redefined principal matrices after changing $L\rightarrow 2L$. Since the initial condition $x(0)$ obeys Eq.~(\ref{eq:boundary2}), one can decompose it as follows:
\begin{align}
x(0) = \sum_{a+t\leq L} c(a,t) B(a,t),
\end{align}
and its time-evolution is given by
\begin{align*}
x(t) = \sum_{a+\tau \leq L} c(a,\tau) B(a,\tau+t).
\end{align*}
Let $(a',t')= R_{2}(x(0))$. Then, from lemma~\ref{lemma_ineq_3dim}, one has
\begin{align*}
W(x(t)) \geq W(B(0,t'+t)).
\end{align*}
Therefore, one has
\begin{align*}
\sum_{t=0}^{L-1} W(x(t)) \geq \sum_{t=0}^{L-1}W (B(0,t+t')) \geq \sum_{t=0}^{L-1}W (B(0,t)) = L^{\mathcal{D}^{(3)}_{p}}
\end{align*}
which completes the proof.

\section{Higher-dimensional fractal code}\label{sec:Ddim}

Finally, we briefly discuss the $D$-dimensional fractal codes for $D>3$. We consider a $D$-dimensional hypercubic lattice with $n = L\times \cdots \times L$ spins with $L=p^{m}$. Each spin is labeled by ``time'' $t$ and ``positions'' $r^{(1)}, \cdots,  r^{(D-1)}$, and we set periodic boundary conditions on all the $D-1$-dimensional surfaces which are parallel to the time axis. The admissible spin configurations of the lattice obey the following constraint:
\begin{align}
x(t+1)_{\textbf{r}} = x(t)_{\textbf{r}} + \sum_{j=1}^{D-1}x(t)_{\tb{r}-\textbf{e}_{j}} \qquad (\mbox{mod $p$})\qquad 0 \leq t \leq L-2 \label{eq:ruleD}
\end{align}
where $\textbf{r}=(r^{(1)}, \cdots,  r^{(D-1)})$, and $\textbf{e}_{j}$ is a unit vector in the $r^{(j)}$ direction. In addition, we limit ourselves to spin configurations which satisfy the following initial condition:
\begin{align}
x(0)_{\textbf{r}}=0 \qquad \mbox{for}\quad \sum_{j=1}^{D-1} r^{(j)} \geq L, \label{eq:boundaryD}
\end{align}
and denote a space of spin configurations specified by the condition above as $C^{(D)}_{p}$. Then, we have the following theorem.

\begin{theorem}[Higher-dimensional fractal code]\label{theorem_main3}
For the codeword space $C^{(D)}_{p}$, let $k$ be the number of encodable spins and $d$ be the code distance of the code. Then, we have
\begin{align}
k= \frac{L(L+1)\cdots(L+D-2)}{(D-1)!}, \qquad d =  L^{\mathcal{D}^{(D)}_{p}}
\end{align}
where 
\begin{align}
\mathcal{D}_{p}^{(D)}= \log \left(\frac{p(p+1)\cdots (p+D-1)}{D!} \right)/\log(p).
\end{align}
\end{theorem}

The fractal dimension $\mathcal{D}^{(D)}_{p}$ goes to $D$: $\mathcal{D}_{p}^{(D)} \rightarrow D$ for $p \rightarrow \infty$. Therefore, the code saturates the bound $k d^{1/D}\leq O(n)$ in Eq.~(\ref{eq:bound}) asymptotically for arbitrary $D$. 

Here, we give a sketch of the proof since it is complicated, but straightforward to obtain the proof. Recall that we have defined two-dimensional principal matrices from one-dimensional principal vectors. We define $(D-1)$-dimensional principal tensors recursively from $(D-2)$-dimensional principal tensors. In particular, a $(D-1)$-dimensional principal tensor $B(\tb{r},t)$ with $(D-2)$-dimensional vector $\tb{r}$ is defined as the time evolution of $B(\tb{r},0)$:
\begin{align}
B(\tb{r},0) =\begin{bmatrix}
B(\tb{r}) \\
0 \\
\vdots \\
0
\end{bmatrix}
\end{align}
where $B(\tb{r})$ is a $(D-2)$-dimensional principal tensor. 

With these independent $(D-1)$-dimensional principal tensors $B(\tb{r},t)$, one can decompose an arbitrary $(D-1)$-dimensional tensor uniquely. One can obtain the following inequality to bound the weight of $(D-1)$-dimensional tensors:
\begin{lemma}
For
\begin{align}
v=\sum_{\tb{r},t}c(\tb{r},t)B(\tb{r},t)\qquad \mbox{where} \qquad c(\tb{r},t)=0\quad \mbox{for all}\quad t+ \sum_{j=1}^{D-2}r_{j} \geq L,
\end{align}
we define the following sets with $R_{D-1}\subseteq \cdots \subseteq R_{1} \subseteq R_{0}$:
\begin{equation}
\begin{split}
R_{0} &= \big\{ (r_{1},\cdots,r_{D-1}) \ :\  c(r_{1},\cdots,r_{D-1}) \not= 0 \big\}\\
R_{a} &= \big\{ (r_{1},\cdots,r_{D-1})\in R_{a-1} \ : \ \sum_{j=a}^{D-1}r_{j} \leq \sum_{j=a}^{D-1}r_{j}' \ \ \mbox{for all}\ \ (r_{1}',\cdots,r_{D-1}')\in R_{a-1} \big\}.
\end{split}
\end{equation}
Then, for $(r_{1}',\cdots,r_{D-1}')\in R_{D-1}$, one has
\begin{align}
W(v) \geq W(B(\tb{0},r_{D-1}'))
\end{align}
\end{lemma}

This bound can be proven recursively by using lemma~\ref{lemma_ineq_3dim}. As a result of this bound, one can easily obtain a lower bound for the weight of arbitrary spin configurations arising in $D$-dimensional fractal codes.

\section{Discussion and open questions}\label{sec:open}

In this paper, we have presented fractal codes which asymptotically saturates the local code bound. There are a number of interesting open questions and future problems, and this section is devoted to discussions and speculations on them. 

\subsection{Open questions on local codes}

An immediate question is whether a local code which ``tightly'' saturates the bound may exist or not. While our discussion was limited to local codes based on the Sierpinski triangle, there are other interesting local codes with various fractal spin configurations. Time evolutions of arbitrary cellular automaton, based on local update rules, can be physically realized as local codes, and lead to fractal spin configurations when update rules are linear~\cite{Wolfram_Text}. The main technical finding in this paper is that the code distance $d$ of the Sierpinski-type fractal code grows with fractal dimensions of the Sierpinski triangle. Whether code distances of generalized fractal codes grow with fractal dimensions of original fractal geometries or not may be an interesting open problem. For instance, the following update rule
\begin{align}
x(t+1)_{r} =  x(t)_{r-1} +x(t)_{r} +x(t)_{r+1} \qquad \mbox{(mod $2$)}
\end{align}
leads to a fractal geometry with a fractal dimension $\frac{\log 1 + \sqrt{5}}{\log 2}$. Then a naturally arising question is whether $d\sim O(L^{\frac{\log 1 + \sqrt{5}}{\log 2}})$ or not. Another important question concerns the necessity of boundary terms for local codes generated by reversible cellular automaton. Finally, one may also consider local codes generated by non-linear cellular automaton. For instance, Wolfram's rule 30 cellular automaton is known to generate pseudo-random spin configurations~\cite{Wolfram_Text} which may achieve $d\sim O(L^{2})$. Also, such a model may be interesting as a toy model of spin glasses without quenched disorder.

While our discussion in this paper is limited to classical error-correcting codes, a similar question for quantum error-correcting codes is also of practical and fundamental importance in quantum information processing. The ``quantum'' information storage capacity for local quantum codes was found in~\cite{Bravyi10}:
\begin{align}
k d^{\frac{2}{D-1}} \leq O(n)
\end{align}
where $d$ is the ``quantum'' code distance. For $D=2$, the Toric code is known to saturate the bound, while the problem of finding a capacity saturating code for $D=3$ is currently open. Recently, there have been significant progresses in systematically studying coding properties of local quantum codes with translation symmetries~\cite{Beni10, Beni10b, Beni11, Haah11, Bombin11b, Kim12}. In particular, a three-dimensional local quantum code with anti-commuting pairs of fractal-like logical operators has been found~\cite{Haah11}. A general framework to extend ``classical'' fractal codes to ``quantum'' fractal codes has been recently obtained along with theoretical tools to compute their code distances~\cite{Beni13}.

\subsection{Physical implementation}

Before discussing the feasibility of physically implementing fractal codes, we need to briefly discuss how bits of information are stored in memory devices that are currently used. First of all, in order to store bits of information securely, one needs to create some stable physical entities with multiple degrees of freedom. Such physical systems may viewed as stable ``spins'' whose sizes are often much larger than sizes of actual single spins or quanta. For instance, in hard disk drives, ferromagnetic materials, physical realizations of repetition codes via local Hamiltonians, are used as stable spins. (Encoding bits of information into actual single spins is technically challenging, and only experimental demonstrations in highly controlled systems are available at this moment). Based on these stable spins, one further encodes bits of information using error-correcting codes by considering stable spins as basic building blocks. These error-correcting codes, used for further encoding bits of information into stable spins, are not necessary supported by local interaction terms since stable spins do not need to be protected by Hamiltonians. For such error-correcting codes without locality, the ultimate bound on information storage capacity is the well-celebrated Shannon bound. It is well known that some error-correcting codes saturate the Shannon bound while admitting efficient decoding of encoded information. Conventional theory of error-correcting codes focuses on the art of encoding based on stable qubits. On the other hand, a problem of finding good local codes focuses on creating stable spins via local Hamiltonians, with a hope of creating stable spins which are more beneficial than ferromagnets, and does not focus on encoding bits of information based on stable spins. 

The biggest obstacle in physically implementing fractal codes is that it involves three-body interactions which may not exist naturally in physical systems. Yet, one may be able to find two-body classical Hamiltonian which leads to the same codeword space as fractal codes via simple magnetic interactions. Another possible approach may be to simulate three-body terms through two-body terms perturbatively~\cite{Jordan08} or non-perturbatively~\cite{Ocko11b} by using quantum Hamiltonians. Finally, it may be possible that Hamiltonians of fractal codes appear as effective Hamiltonians of some known interacting spin systems. 

Encoding and decoding bits of information in an efficient and physically implementable way is also an important problem from a practical viewpoint. From coding theory perspective, practical encoding and decoding algorithm may exist for local codes since local codes can be viewed as low-density parity-check codes (LDPCs)~\cite{Gallager62}. A particularly interesting approach, motivated by physical arguments, is the so-called renormalization group (RG) decoding algorithm~\cite{Duclos-Cianci10}. These algorithms, however, require some ``computations'' to write and read out bits of information. This is in a strong contrast with the fact that, for a ferromagnet, one can easily write and read a single bit of information just by adding external magnetic fields and by measuring the total magnetization, without any computations. One needs to find encoding and decoding algorithm for fractal codes which are physically motivated and are implementable without any non-local computations.

\subsection{As a many-body spin system}

Studies on physical properties of fractal codes may also be of fundamental interest in condensed matter physics community. Searching for novel local codes is fundamentally akin to searching for novel quantum phases as local codes can be viewed as representatives of quantum phases with mass gap~\cite{Beni10b}. Most of conventional many-body spin systems, such as a ferromagnet, are known to have continuous scale symmetries where ground states look exactly the same even after changing the length scale of the system. This observation led to the development of the renormalization group theory for classifying quantum phases arising in many-body systems. In this light, fractal codes are unconventional since their ground states do not have continuous scale symmetries, but have only discrete scale symmetries where ground states of fractal codes with $p$-dimensional spins look the same only under the scale transformations by powers of $p$. Clearly, systems with discrete scale symmetries are beyond descriptions of topological field theory, and searches for their effective theories may be an interesting future problem. Many-body physics with discrete scale symmetries is a largely uncharted research area except some pioneering works~\cite{Efimov70, Wilson71}. 

While our discussion was limited to the ground state properties of fractal codes, properties of quasi-particle excitations in fractal codes are also interesting. It should be noted that thermal relaxation dynamics of a fractal code with $p=2$ was studied in~\cite{Newman99} more than a decade ago where a fractal spin model was originally proposed as a toy model which may exhibit spin-glass like relaxation dynamics even without quenched disorder. In particular, it was shown that different ground states of fractal spin systems are separated by energy barriers which grow logarithmically with respect to the system size. 

\subsection{Information storage capacity in other physical systems}

Another intriguing question concerns a connection between the Bekenstein bound and the local code bound. The Bekenstein bound is a quantum mechanical bound on the degree of freedom on a finite physical space, resulting from the uncertainty principle. When the Bekenstein bound is applied to systems with gravity, one can derive the area law for black hole entropies by requiring that the size of an object does not exceed the Schwarzschild radius and can find that entropies are upper bounded roughly by $A/\ell_{p}^{2}$ where $A$ is the surface are of an object and $\ell_{p}$ is the Planck length. This observation led to the holographic principle of black holes which essentially states that physical states of black holes can be determined completely by the surface of black holes. 

As for the local code bound, it is interesting to observe that an area law naturally arises in fractal codes; the number of encoded bits $k$ is area-like with $k\sim O(L^{D-1})$, while the code distance $d$ is asymptotically volume-like with $d \sim O(L^{D-\epsilon})$. In deriving the local code bound, only the locality of interaction terms on a discretized space is assumed. It is interesting to note that, the area-law arising in fractal codes can be derived from purely classical calculations while derivations of black hole area-law and entanglement entropy area-law both crucially require quantum mechanics. A construction of fractal codes is, in some sense, rooted on the holographic principle where ground states can be uniquely specified by spin values on the surface. However, a connection between fractal codes and black holes has not been established. 

In the present paper, our discussion has been constrained to the following three conditions; a) static Hamiltonian, b) local interactions, and c) being frustration-free. There are a large number of pioneering works that addressed a similar question without above three constraints. A problem of encoding bits of information into dynamically evolving systems has been actively addressed in studies of neural networks. For instance, the Hopfield model of neural network, based on the Hebb rule, is capable of reliably storing a large amount of information which easily breaks the local code bound if non-local couplings are allowed. In~\cite{Gacs01}, G\'acs presented a model of one-dimensional locally coupled dynamical spin systems that is capable of storing one bit per site, and is still robust against small, but finite amount of noises. This remarkable result by G\'acs implies that dynamical systems are more powerful than static systems in terms of information storage capacity. While we have limited our considerations only to frustration-free spin systems, there are a large number of interesting frustrated spin systems including spin glasses and anti-ferromagnets which may be useful in storing bits of information. Relations between spin glass systems and classical error-correcting codes have been actively investigated where some classes of spin glasses with non-local couplings are known to saturate the Shannon bound asymptotically~\cite{Sourlas89}. 

\section*{Acknowledgments}

I thank Eddie Farhi and Peter Shor for support at MIT. I thank Sergey Bravyi, Patrick Hayden, Masahito Ueda and John Preskill for comments and discussion. This work is supported in part by DOE Grant No. DE-FG02-05ER41360 and by Nakajima Foundation.

\appendix

\section{Review of coding theory}\label{sec:intro}

We give a brief review of theory of classical error-correcting codes in the context of spin physics. We also give a derivation of the local code bound, following~\cite{Bravyi10}. A precise definition of frustration-free classical local Hamiltonians is also given here. \\

\tb{Local code:}
Consider a $D$-dimensional hyper-cubic lattice of $L\times \cdots \times L$ spins whose spin values are $0$ or $1$. A classical local Hamiltonian $H$ can be written in the following form
\begin{align}
H = \sum_{a=1}^{m} \Pi_{a}
\end{align}
where interaction terms $\Pi_{a}$ are supported locally inside some finite regions of $\omega \times \cdots \times \omega$ spins. Here, $\omega$ is referred to as a \emph{range} of interactions. 

A Hamiltonian is said to be \emph{frustration-free} when a energy ground state can be obtained by minimizing each interaction term $\Pi_{a}$ independently. Without loss of generality, we assume that the smallest value of $\Pi_{a}$ is zero for all $a$; $\Pi_{a}\geq 0$. Then, a ground state $\tb{s}$ of a frustration-free Hamiltonian $H$ satisfies
\begin{align}
\Pi_{a}(\bf{s}) = 0, \qquad \mbox{for all $a$}.
\end{align}
We call such classical frustration-free Hamiltonians \emph{local codes}. Ground states of local codes can be viewed as binary strings, and form the codeword space (the ground space) $\mathcal{C}$:
\begin{align}
\mathcal{C} = \{ \tb{s} : \Pi_{a}(\tb{s})=0, \quad \forall a \}.
\end{align}
The number of encoded logical bits is $k=\log_{2}\dim \mathcal{C}$, and there are $2^{k}$ degenerate ground states in a local code. \\

\tb{Code distance:}
Let us estimate how reliably bits of information can be stored in the presence of errors. Suppose one encodes bits of information into a ground state $\tb{s}_{0}$. Then, consider an error which flips some spins, giving rise to a spin configuration denoted by $\tb{s}_{error} \not= \tb{s}_{0}$. If the number of flipped spins is small, $\tb{s}_{error}$ will be still close to the original ground state $\tb{s}_{0}$, so one may be able to recover the encoded information. If the number of flipped spins is large, $\tb{s}_{error}$ may be closer to other ground states $\tb{s}_{j}$ ($j\not=0$), so one may not be able to recover the encoded information. Based on this observation, it is convenient to introduce the Hamming distance between two binary strings $\tb{s}$ and $\tb{s}'$ which is the number of different spin values in $\tb{s}$ and $\tb{s}'$, corresponding to the weight of $\tb{s}+\tb{s}'$ (mod $2$). The Hamming distance is the number of spin flips necessary to change from $\tb{s}$ to $\tb{s}'$, and the code distance $d$ is defined as the minimal Hamming distance between all the possible pairs of ground states:
\begin{align}
d = \min w(\tb{s}+\tb{s}'),\quad \forall \tb{s},\tb{s}'\in \mathcal{C}
\end{align}
where $\tb{s}\not=\tb{s}'$.\\

\tb{Singleton bound:}
To derive the local code bound, it is convenient to recall a certain fundamental upper bound on classical error-correcting codes, called the Singleton bound:
\begin{align}
n-d+1 \geq k.
\end{align}
We emphasize that this bound holds without assuming the geometric locality of $\Pi_{a}$ (i.e. for an arbitrary interaction range $\omega$). The Singleton bound can be proven by considering a bi-partition of the entire system into $A$ and $B$ where $B$ consists of $d-1$ spins and $A$ consists of $n-d+1$ spins. Suppose that there exist two ground states $\tb{s}$ and $\tb{s}'$ whose spin values inside $A$ are exactly the same: $\tb{s}|_{A}=\tb{s}'|_{A}$. If $\tb{s}$ and $\tb{s}'$ are different ground states, one can obtain $\tb{s}'$ from $\tb{s}$ just by flipping spins only inside $B$. But this contradicts with the fact that the code distance is $d$ while the number of spins inside $B$ is $d-1$. So, $\tb{s}$ and $\tb{s}'$ must be the same ground state. This implies that, if $\tb{s} \not=\tb{s}'$, their spin values in $A$ must be different: $\tb{s}|_{A}\not=\tb{s}'|_{A}$. Therefore, the number of logical bits $k$ is upper bounded by the number of spins in $A$, which is $n-d+1$.\\

\tb{Local code bound:}
One can extend the Singleton bound for local codes by imposing geometric locality on interaction terms $\Pi_{a}$. The key idea in proving the Singleton bound is to remove a region $B$ whose volume is smaller than $d$, and upper-bound the number of logical bits $k$ by the number of spins in $A$. To prove the local code bound, we think of a bi-partition into $A$ and $B$ where $B$ consists of hyper-cubic blocks $B=B_{1}\cdots B_{m}$ whose sizes are smaller than $d$ and their separation is at least $\omega$ so that interaction terms $\Pi_{a}$ may overlap with at most one block at the same time. We think of two ground states $\tb{s}$ and $\tb{s}'$ such that $\tb{s}|_{A}=\tb{s}'|_{A}$. Then, one can obtain $\tb{s}'$ from $\tb{s}$ by flipping spins inside $B=B_{1}\cdots B_{m}$ where we denote sets of spins inside $B_{j}$ which are to be flipped by $E_{j}\subseteq B_{j}$. Let us consider a ground state $\tb{s}_{1}$ which can be obtained from $\tb{s}$ by flipping spins in $E_{1}$. Then, due to the locality of interaction terms $\Pi_{a}$, $\tb{s}_{1}$ is also a ground state. Since the size of $B_{1}$ is smaller than $d$, $E_{1}$ must be a null set. Similarly, one has $E_{j}=0$ for all $j$ and thus, $\tb{s}=\tb{s}'$. This implies that if $\tb{s} \not=\tb{s}'$, their spin values in $A$ must be different: $\tb{s}|_{A}\not=\tb{s}'|_{A}$, and one has
\begin{align}
v_{A} \geq k
\end{align}
where $v_{A}$ is the number of spins in $A$. 

One needs to find an upper bound on $v_{A}$ by removing as many blocks $B_{j}$ as possible while keeping the separations between blocks $B_{j}$ to be at least $\omega$. Let us think of cubic regions $B_{j}$ whose linear length is of order $\ell \equiv d^{1/D}$. Then, the number of blocks which can removed is of order
\begin{align}
\left( \frac{L}{\omega + \ell}\right)^{D}.
\end{align}
Therefore, $v_{A}$ is upper bounded roughly by
\begin{align}
v_{A}\leq L^{D} - \ell^{D} \cdot \left( \frac{L}{\omega + \ell}\right)^{D}.
\end{align} 
This leads to 
\begin{align}
n \geq k \frac{(\omega + \ell)^{D} }{  (\ell + \omega)^{D} - \ell^{D}   }, \qquad \ell = d^{1/D}.
\end{align}
This leads to the local code bound $kd^{1/D}\leq O(n)$.

\section{Proofs of some lemmas}\label{sec:proof}

In this appendix, we give proofs of some lemmas used in the main discussion.

\subsection{Proof of lemma~\ref{lemma_ineq}}\label{appendix:A2}

The proof of lemma~\ref{lemma_ineq} consists of several steps.  \\

\tb{Inverse matrices:}
We begin by finding the inverse matrix $\tb{B}^{-1}$ of the Pascal matrix $\tb{B}$:

\begin{lemma}\label{lemma_inverse}
The inverse matrix is given by:
\begin{align}
B^{-1}(t)_{r}= B(L-1-r)_{L-1-t}
\end{align}
where $B^{-1}(t)_{r}$ represents an entry of $\tb{B}^{-1}$ at $(t,r)$. In particular, its entries are given by
\begin{align}
B^{-1}(t)_{r} = {}_{L-1-r} C_{L-1-t} = (-1)^{t+r}{}_{t} C_{r} =  (-1)^{t+r}B(t)_{r}.
\end{align}
\end{lemma}

\begin{figure}[htb!]
\centering
\includegraphics[width=0.45\linewidth]{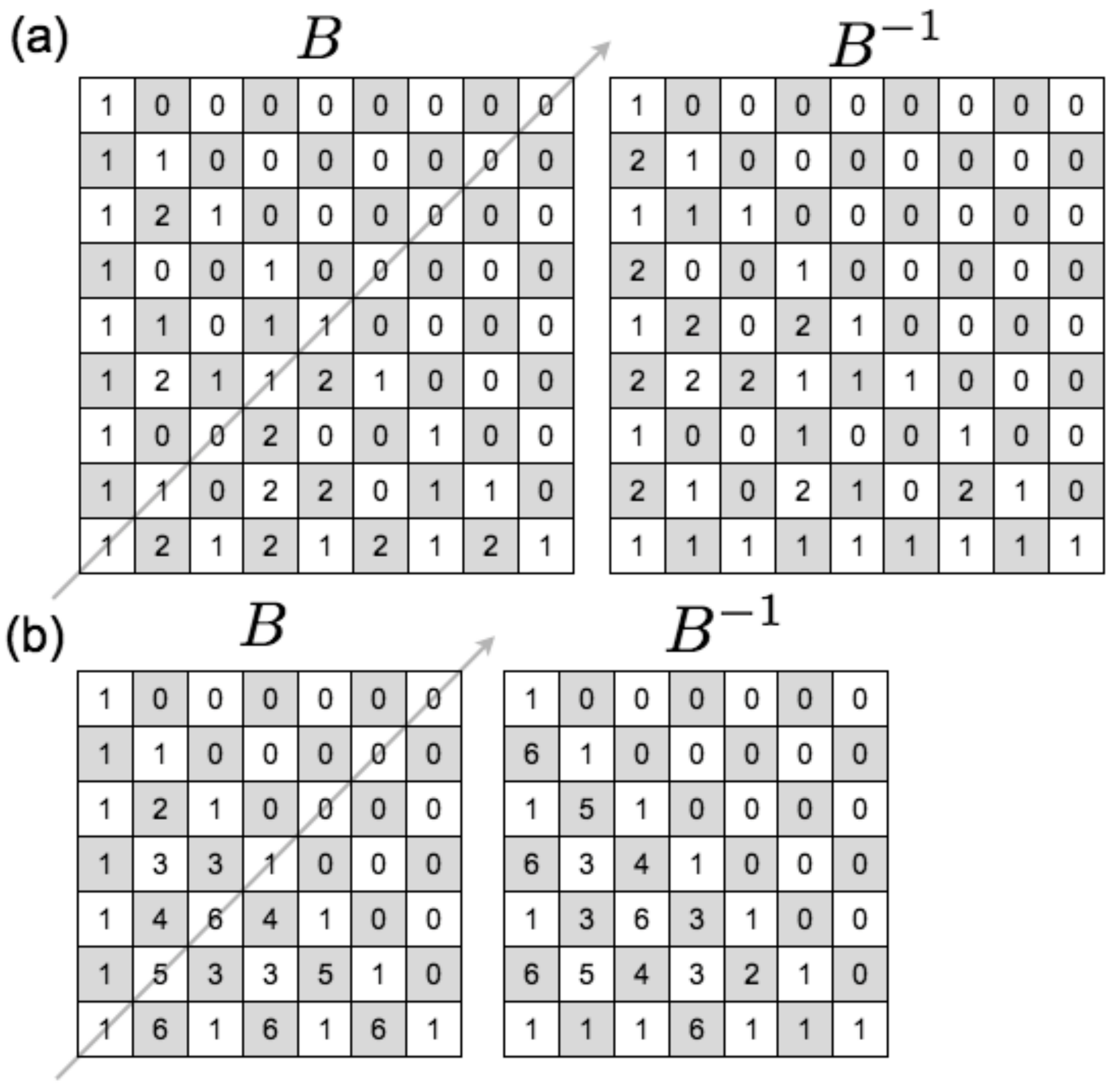}
\caption{Examples of the Pascal matrices and its inverse matrices. Only the shaded regions with $t+r=\mbox{odd}$ may have different entries in $\tb{B}$ and $\tb{B}^{-1}$. The inverse matrices can be obtained by reflecting the original matrices along the arrows shown above. (a) $p=3$ and $m=2$. (b) $p=7$ and $m=1$.
} 
\label{fig_inverse}
\end{figure}

Examples of inverse matrices are shown in Fig.~\ref{fig_inverse}. 

\begin{proof}
Since $B^{-1}(t)_{r}= B(L-1-r)_{L-1-t}$, we have
\begin{align*}
B^{-1}(t)_{r} = {}_{L-1-r} C_{L-1-t} &= \frac{(p^{m}-t)\cdots(p^{m}-1-r)}{(t-r)!}\\
                       &= \frac{(-t)(-t+1)\cdots(-1-r) }{(t-r)!} \\
                       &= \frac{(-1)^{t-r} (r+1) \cdots (t-1)t}{(t-r)!}\\
                       &= \frac{(-1)^{t+r} t!}{r!(t-r)!}\\
                       &= (-1)^{t+r}{}_{t} C_{r} 
\end{align*}
for $t \geq r$ where all the calculations are carried out modulo $p$. It is straightforward to see that $\tb{B}\cdot \tb{B}^{-1}=\tb{I}$ with some calculations.
\end{proof}

The following lemma is useful in finding the power $\tb{B}^{c}$ of the Pascal matrix $\tb{B}$:

\begin{lemma}\label{lemma_power}
A matrix $\tb{B}^{c}$ is generated by the following modified rule
\begin{align}
x(t+1)_{r} = x(t)_{r-1} + cx(t)_{r} \qquad (\mbox{mod $p$})\qquad 0 \leq t \leq L-2 
\end{align}
with an initial condition $x(0)=(1,0,\cdots)$.
\end{lemma}

\begin{figure}[htb!]
\centering
\includegraphics[width=0.50\linewidth]{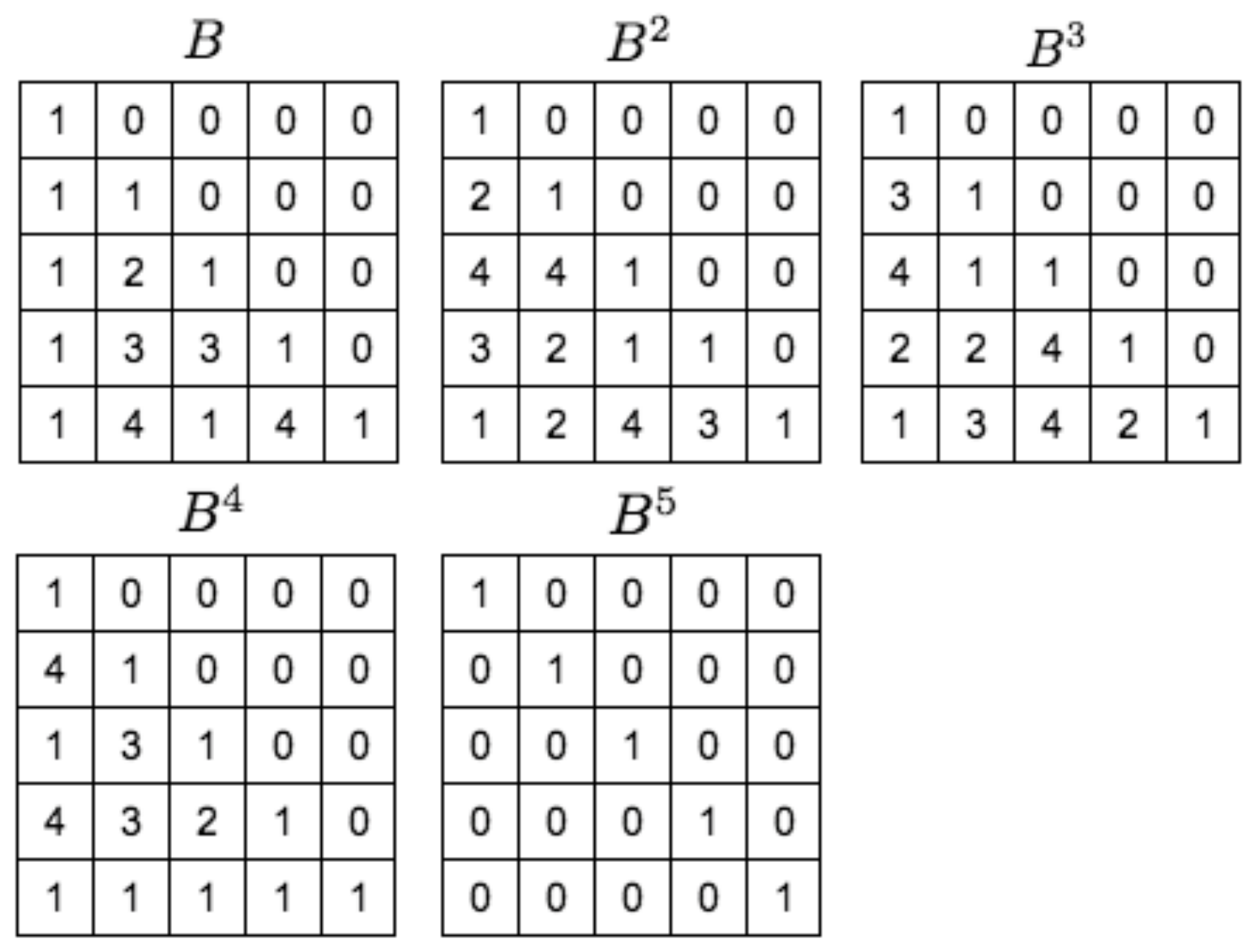}
\caption{Examples of powers of the Pascal matrix $\tb{B}$ for $p=5$.
} 
\label{fig_inverse2}
\end{figure}

By modifying the local rule for spin configurations, one can obtain powers of the Pascal matrix. An example is shown in Fig.~\ref{fig_inverse2}. Here, we notice that $\tb{B}^{p}=\tb{I}$ since the update rule is reduced to 
\begin{align}
x(t+1)_{r} = x(t)_{r-1}  \qquad (\mbox{mod $p$})\qquad 0 \leq t \leq L-2.
\end{align}
Also, we notice that $\tb{B}^{p-1}=\tb{B}^{-1}$ from $\tb{B}^{p}=\tb{I}$.

\begin{proof}
For simplicity of discussion, we only prove the lemma for $\tb{B}^{2}$. Since 
\begin{align*}
B(t)_{r} = B(t-1)_{r} + B(t-1)_{r-1},
\end{align*}
we have
\begin{align*}
B^{2}(t)_{r} &= \sum_{a} B(t)_{a}B(a)_{r} \\
                      &= \sum_{a}(B(t-1)_{a} + B(t-1)_{a-1})B(a)_{r} \\
                      &= B^{2}(t-1)_{r} + \sum_{a} B(t-1)_{a-1}B(a)_{r} \\
                      &= B^{2}(t-1)_{r} + \sum_{a} B(t-1)_{a}B(a+1)_{r} \\
                      &= B^{2}(t-1)_{r} + \sum_{a} (B(t-1)_{a}B(a)_{r} + B(a)_{r-1}) \\
                      &= B^{2}(t-1)_{r} + B^{2}(t-1)_{r} + B^{2}(t-1)_{r-1} = 2 B^{2}(t-1)_{r} + B^{2}(t-1)_{r-1} 
\end{align*}
where $B(t)_{a}=0$ for $t<0$, and $B(t)_{a}=B(t)_{a+L}$ if $a<0$. Then, we notice that entries of $\tb{B}^{2}$ obeys the rule in Eq.~(\ref{eq:rule2}) for $c=2$. Therefore, $\tb{B}^{2}$ can be generated from the rule for $c=2$. A similar discussion leads to the proof for an arbitrary $c$. 
\end{proof}

\tb{Submatrices of the Pascal matrix:} 
The the Pascal matrix $\tb{B}$ is invertible since principal vectors $B(t)$ are pairwise independent. Similar properties holds for \emph{submatrices} of $\tb{B}$. We denote the Pascal matrix $\tb{B}$ for $m=1$ as $\tb{B}^{(1)}$:
\begin{align}
\tb{B}^{(1)} = 
\begin{bmatrix}
{}_{0} C_{0},& {}_{0} C_{1},& {}_{0} C_{2},& \cdots & {}_{0} C_{p-1}\\
{}_{1} C_{0},& {}_{1} C_{1},& {}_{1} C_{2},& \cdots & {}_{1} C_{p-1}\\
\vdots          &  \vdots         &    \vdots       & \vdots & \vdots \\
{}_{p-1} C_{0},& {}_{p-1} C_{1},& {}_{p-1} C_{2},& \cdots & {}_{p-1} C_{p-1}
\end{bmatrix}\qquad \mbox{(mod $p$)}
\end{align}
which is a $p \times p$ matrix. Then, for submatrices of $\tb{B}^{(1)}$, we have the following lemma:

\begin{lemma}\label{lemma_submatrix}
Consider the following submatrix of $\tb{B}^{(1)}$:
\begin{align}
\tb{A} =
\begin{bmatrix}
{}_{x_{0}} C_{0},& {}_{x_{0}} C_{1},& {}_{x_{0}} C_{2},& \cdots & {}_{x_{0}} C_{a}\\
{}_{x_{1}} C_{0},& {}_{x_{1}} C_{1},& {}_{x_{1}} C_{2},& \cdots & {}_{x_{1}} C_{a}\\
\vdots          &  \vdots         &    \vdots       & \vdots & \vdots \\
{}_{x_{a}} C_{0},& {}_{x_{a}} C_{1},& {}_{x_{a}} C_{2},& \cdots & {}_{x_{a}} C_{a}
\end{bmatrix}
\end{align}
where $a < p$ and $0 \leq x_{0}< x_{1} <\cdots < x_{a} < p$. Then, $\tb{A}$ always has an inverse matrix $\tb{A}^{-1}$. Similarly, consider the following submatrix of $\tb{B}^{(1)}$:
\begin{align}
\tb{A'} =
\begin{bmatrix}
{}_{p-a-1} C_{y_{0}},& {}_{p-a-1} C_{y_{1}},& {}_{p-a-1} C_{y_{2}},& \cdots & {}_{p-a-1} C_{y_{a}}\\
\vdots          &  \vdots         &    \vdots       & \vdots & \vdots \\
{}_{p-2} C_{y_{0}},& {}_{p-2} C_{y_{1}},& {}_{p-2} C_{y_{2}},& \cdots & {}_{p-2} C_{y_{a}}\\
{}_{p-1} C_{y_{0}},& {}_{p-1} C_{y_{1}},& {}_{p-1} C_{y_{2}},& \cdots & {}_{p-1} C_{y_{a}}
\end{bmatrix}
\end{align}
where $0\leq y_{0} < y_{1} < \cdots < y_{a} < p$. Then, $\tb{A'}$ always has an inverse matrix $(\tb{A'})^{-1}$.
\end{lemma}

\begin{figure}[htb!]
\centering
\includegraphics[width=0.75\linewidth]{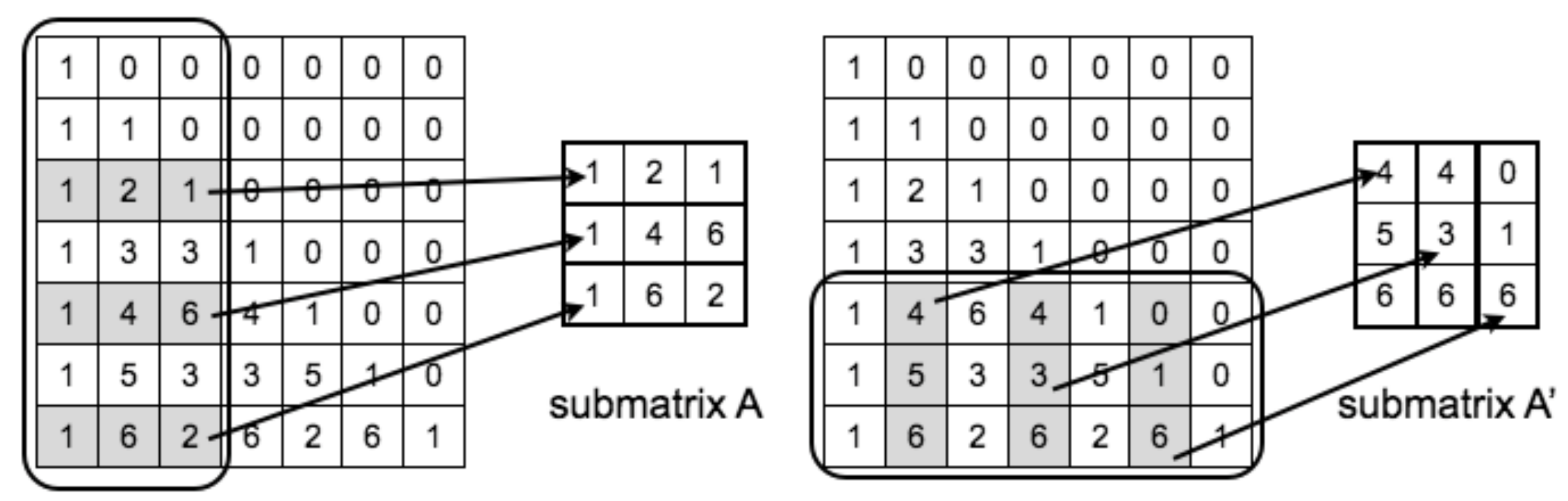}
\caption{Submatrices for $p=7$.
} 
\label{fig_submatrix}
\end{figure}

The construction of $\tb{A}$ goes as follows. First, we choose an $p \times a$ submatrix from $\tb{B}^{(1)}$ on the left hand side of $\tb{B}^{(1)}$ ($a<p$). Then, we pick up $a$ raws to create an $a\times a$ matrix $\tb{A}$. Similarly, to construct $\tb{A'}$, we choose an $a \times p$ submatrix on the bottom of $\tb{B}^{(1)}$, and pick up $a$ columns to create an $a\times a$ matrix $\tb{A'}$. Examples of such constructions of submatrices are shown in Fig.~\ref{fig_submatrix}.

It is worth looking at examples. Consider the case where $p=5$. Then, we have
\begin{align}
\tb{B}^{(1)}=
\begin{bmatrix}
1 ,& 0 ,& 0 ,& 0 ,& 0 \\
1 ,& 1 ,& 0 ,& 0 ,& 0 \\
1 ,& 2 ,& 1 ,& 0 ,& 0 \\
1 ,& 3 ,& 3 ,& 1 ,& 0 \\
1 ,& 4 ,& 1 ,& 4 ,& 1 
\end{bmatrix}.
\end{align}
For $a=2$ and $x_{0}=2$, $x_{1}=3$ and $x_{2}=4$, we have
\begin{align}
\tb{A}=
\begin{bmatrix}
1 ,& 2 ,& 1  \\
1 ,& 3 ,& 3  \\
1 ,& 4 ,& 1  
\end{bmatrix}.
\end{align}
Since three vectors $(1,2,1)$, $(1,3,3)$ and $(1,4,1)$ are independent (mod $5$), $\tb{A}$ is invertible. For $a=2$ and $y_{0}=1$, $y_{1}=2$ and $y_{2}=3$, we have 
\begin{align}
\tb{A'}=
\begin{bmatrix}
2 ,& 1 ,& 0  \\
3 ,& 3 ,& 1  \\
4 ,& 1 ,& 4  
\end{bmatrix}
\end{align}
which is also invertible.

\begin{proof} For simplicity of discussion, we present a proof only for $\tb{A}$. First, recall that the Vandermonde matrix $\tb{M}$ has the following well-known property:
\begin{align}
\tb{M} = \begin{bmatrix}
1 ,& x_{0} ,& x_{0}^{2},& \cdots ,& x_{0}^{a} \\
1 ,& x_{1} ,& x_{1}^{2},& \cdots ,& x_{1}^{a} \\ 
\vdots & \vdots & \ddots & \vdots \\
1 ,& x_{a} ,& x_{a}^{2},& \cdots ,& x_{a}^{a} \\
\end{bmatrix},\qquad
\det(\tb{M})= \prod_{0 \leq i< j \leq a } (x_{j}-x_{i}).
\end{align}
Therefore, the following vectors are independent when $x_{i}\not=x_{j}$ for all $i$ and $j$:
\begin{align}
(1,1,\cdots,1), (x_{0},x_{1},\cdots,x_{a}) \quad \cdots \quad (x_{0}^{a},x_{1}^{a},\cdots,x_{a}^{a}).\label{eq:indep}
\end{align}

Now, let us consider the Vandermonde matrix modulo $p$. When $0 \leq x_{0}< x_{1} <\cdots < x_{a} < p$, the vectors above are independent modulo $p$ since the determinant of $\tb{M}$ computed modulo $p$ is nonzero. Notice that our goal to prove that the following vectors are independent modulo $p$:
\begin{align}
({}_{x_{0}} C_{0},{}_{x_{1}} C_{0},\cdots,{}_{x_{a}} C_{0}),({}_{x_{0}} C_{1},{}_{x_{1}} C_{1},\cdots,{}_{x_{a}} C_{1}) \quad \cdots \quad ({}_{x_{0}} C_{a},{}_{x_{1}} C_{a},\cdots,{}_{x_{a}} C_{a}). \label{eq:indep2}
\end{align}
This is immediate since vectors in Eq.~(\ref{eq:indep2}) can be created by adding vectors in Eq.~(\ref{eq:indep}). Therefore, $\tb{A}$ is invertible. A similar proof works for $\tb{A'}$ by using lemma~\ref{lemma_inverse}.
\end{proof}

\tb{Proof of lemma~\ref{lemma_ineq}:} 
Finally, we prove lemma~\ref{lemma_ineq}. Due to the self-similar structures of the Pascal matrix $\tb{B}$, it is sufficient to prove the lemma for $m=1$. Consider a decomposition
\begin{align*}
v=\sum_{t} c(t)B(t)
\end{align*}
where $t_{min}$ is the minimal integer such that $c(t)\not=0$. Then, the goal is to prove 
\begin{align*}
W(v) \geq W(B(t_{min})).
\end{align*}

We list all the integers $r$ such that
\begin{align*}
B(t_{min})_{r}\not=0 \qquad \mbox{and} \qquad v_{r}=0
\end{align*}
and denote them as $r_{1},\cdots,r_{a}$. Then, the number of non-zero entries of $v_{r}$ for $r\leq t_{min}$ is $W(B(t_{min}))-a$. Next, we list all the integers $r$ such that 
\begin{align*}
B(t_{min})_{r}=0 \qquad \mbox{and} \qquad v_{r}=0
\end{align*}
and denote them as $r_{1}',\cdots,r_{b}'$. Then, the number of non-zero entries of $v_{r}$ for $t_{min}<r$ is $(p-1-t_{min})-b$. Then, we have
\begin{align*}
W(v) = W(B(t_{min})) - a + (p-1-t_{min}) -b 
\end{align*}
from a simple counting argument. Therefore, it suffices to prove that $a+b \leq p-1-t_{min}$.

We next consider constrains on coefficients $c(t)$:
\begin{align*}
v_{r}=\sum_{t=0}^{p-1} c(t)B(t)_{r} =0 \qquad \mbox{for}\quad r=r_{1},\cdots, r_{a}, r_{1}',\cdots, r_{b}'.
\end{align*}
Recall that $t_{min}$ is the minimal $t$ such that $c(t)\not=0$ and $c(t)=0$ for $t<t_{min}$. Then, the above constraints can be concisely represented as follows:
\begin{align*}
(\tb{A'})^{T}\cdot 
\begin{bmatrix}
c(t_{min}) \\
c(t_{min}+1) \\
\vdots \\
c(p-1)
\end{bmatrix} = 0
\end{align*}
where
\begin{align*}
\tb{A'} = \begin{bmatrix}
B(t_{min})_{r_{1}}   ,& \cdots, & B(t_{min})_{r_{a}}   ,& B(t_{min})_{r_{1}'}   ,& \cdots, & B(t_{min})_{r_{b}'}    \\
B(t_{min}+1)_{r_{1}} ,& \cdots, & B(t_{min}+1)_{r_{a}} ,& B(t_{min}+1)_{r_{1}'} ,& \cdots, & B(t_{min}+1)_{r_{b}'}  \\
\vdots                & \vdots  & \vdots                & \vdots                & \vdots  & \vdots                \\
B(p-1)_{r_{1}}       ,& \cdots, & B(p-1)_{r_{a}}       ,& B(p-1)_{r_{1}'}       ,& \cdots, & B(p-1)_{r_{b}'}       
\end{bmatrix}.
\end{align*}
Here, $\tb{A'}$ is a $(p-t_{min})\times (a+b)$ matrix. Notice that the rank of $\tb{A'}$ is $p-t_{min}$ when $a + b \geq p-t_{min}$ due to lemma~\ref{lemma_submatrix}. In such cases, we have $c(t)=0$ for all $t$ which leads to a contradiction. Therefore, $a + b < p-t_{min}$. This leads to $W(v) \geq W(B_{min})$, and completes the proof of lemma~\ref{lemma_ineq}.

\subsection{Proof of lemma~\ref{lemma_ineq_3dim}}\label{appendix:A4}

Lemma~\ref{lemma_ineq_3dim} is an inequality concerning the weights of principal matrices. We begin by proving the following lemma.

\begin{lemma}\label{lemma_R1}
Consider a matrix 
\begin{align}
v = \sum_{a,t}c(a,t)B(a,t)
\end{align}
where $(c,t)=0$ for $a+t\geq L$. Let $v^{*}$ be
\begin{align}
v^{*}=\sum_{(a,t)\in R_{1}(v) } c(a,t)B(a,t).
\end{align}
Then, one has
\begin{align}
W(v) \geq W(v^{*}). 
\end{align}
\end{lemma}

\begin{proof}
We decompose $v$ as follows:
\begin{align}
v = \sum_{a+t=0} c(a,t)B(a,t) + \sum_{a+t=1} c(a,t)B(a,t) + \sum_{a+t=2} c(a,t)B(a,t) +\cdots.
\end{align}
In particular, we set
\begin{align}
v = \sum_{\delta} v^{(\delta)},\qquad v^{(\delta)}=\sum_{a+t=\delta} c(a,t)B(a,t).
\end{align}

For simplicity of discussion, we consider the case where
\begin{align*}
v = v^{(\delta)} + v^{(\delta+1)}.
\end{align*}
$v^{(\delta)}$ can be represented as follows:
\begin{align*}
v^{(\delta+1)} = \begin{bmatrix}
d(\delta)\cdot B(\delta)\\
d(\delta-1)\cdot B(\delta-1)\\
\vdots \\
d(0)\cdot B(0) \\
0 \\
\vdots  
\end{bmatrix}
\end{align*}
where $d(0),\cdots, d(\delta)$ are some integers, and $v^{(\delta+1)}$ can be represented as follows:
\begin{align*}
v^{(\delta+1)} = \begin{bmatrix}
d'(\delta+1)\cdot B(\delta+1)\\
d'(\delta)\cdot B(\delta)\\
\vdots \\
d'(0)\cdot B(0) \\
0 \\
\vdots  
\end{bmatrix}
\end{align*}
where $d'(0),\cdots, d'(\delta+1)$ are some integers. Then, we have
\begin{align*}
v^{(\delta)} + v^{(\delta+1)} = \begin{bmatrix}
d'(\delta+1)\cdot B(\delta+1) + d(\delta)\cdot B(\delta)\\
d'(\delta)\cdot B(\delta)+ d(\delta-1)\cdot B(\delta-1) \\
\vdots \\
d'(1)\cdot B(1) +  d(0)\cdot B(0) \\
d'(0)\cdot B(0) \\
0 \\
\vdots  
\end{bmatrix}
\end{align*}
Then, due to lemma~\ref{lemma_ineq}, we have
\begin{align*}
W(v) \geq W(v^{(\delta)}).
\end{align*}
The discussion above can be easily extended to general cases. This completes the proof.
\end{proof}

Next, it is convenient to consider the transposes of principal matrices. Let us begin with an example for $p=2$ and $m=2$:
\begin{align*}
B(1,1) = \begin{bmatrix}
B(2) \\
B(1) \\
0 \\
0 
\end{bmatrix}=
\begin{bmatrix}
1 ,& 0 ,& 1 ,& 0 \\
1 ,& 1 ,& 0 ,& 0 \\
0 ,& 0 ,& 0 ,& 0 \\
0 ,& 0 ,& 0 ,& 0 
\end{bmatrix}.
\end{align*}
Then, its transpose is
\begin{align*}
B(1,1)^{T} = 
\begin{bmatrix}
1 ,& 1 ,& 0 ,& 0 \\
0 ,& 1 ,& 0 ,& 0 \\
1 ,& 0 ,& 0 ,& 0 \\
0 ,& 0 ,& 0 ,& 0 
\end{bmatrix}
=
\begin{bmatrix}
B(1) \\
B(0) + B(1) \\
B(0) \\
0 
\end{bmatrix}.
\end{align*}

Here, we apply lemma~\ref{lemma_R1} to the transpose $B(1,1)^{T}$:
\begin{align*}
B(1,1)^{T} = (B(1,1)^{T})^{(1)} + (B(1,1)^{T})^{(2)}
\end{align*}
where
\begin{align*}
(B(1,1)^{T})^{(1)} =
\begin{bmatrix}
B(1) \\
B(0) \\
0 \\
0 
\end{bmatrix},\qquad (B(1,1)^{T})^{(2)} = \begin{bmatrix}
0 \\
B(1) \\
B(0) \\
0 
\end{bmatrix}.
\end{align*}
Then, one has
\begin{align*}
W(B(1,1)) \geq W((B(1,1)^{T})^{(1)}  ).
\end{align*}
By noticing 
\begin{align*}
(B(1,1)^{T})^{(1)} = B(0,1)=
\begin{bmatrix}
B(1) \\
B(0) \\
0 \\
0 
\end{bmatrix},
\end{align*}
one has 
\begin{align*}
W(B(1,1)) \geq W(B(0,1)).
\end{align*}

As the example above shows, by considering the transpose of principal matrices, one can further lower bound on the weight of principal matrices. For this purpose, the following lemma is particularly useful.

\begin{lemma}\label{lemma_transpose}
Consider a transpose $B(a,t)^{T}$ of a principal matrix with $a+t <L$. For a decomposition of $B(a,t)^{T}$ in terms of principal matrices:
\begin{align}
B(a,t)^{T} = \sum_{a',t'} c(a',t')B(a',t,),
\end{align}
one has
\begin{align}
R_{1}(B(a,t)^{T}) = \{(0,t)\}.
\end{align}
\end{lemma}

The proof of the lemma~\ref{lemma_transpose} is immediate by noticing the following fact:

\begin{fact}
The principal matrices $B(0,t)$ are symmetric under the transpose:
\begin{align}
B(0,t)^{T}=B(0,t).
\end{align}
The principal matrix $B(a,t)$ for $a\not=0$ is given by
\begin{align}
B(a,t) = \sum_{x} T_{r^{(1)}}^{x-1} (B(a)_{x} \cdot B(0,t) ),
\end{align}
and its transpose is given by
\begin{align}
B(a,t)^{T} = \sum_{y} T_{r^{(2)}}^{y-1} (B(a)_{x} \cdot B(0,t) ),
\end{align}
\end{fact}

We finally prove lemma~\ref{lemma_ineq}. For a decomposition
\begin{align*}
v=\sum_{a,t} c(a,t)B(a,t),
\end{align*}
from lemma~\ref{lemma_R1}, one has
\begin{align*}
W(v)\geq W(v^{*}). 
\end{align*}
Here, we consider the transpose of $v^{*}$. Then, for $(a',t')\in R_{2}(v)$, from lemma~\ref{lemma_transpose}, one has
\begin{align*}
R_{1}((v^{*})^{T}) = (0,t').
\end{align*}
Therefore, we have
\begin{align*}
W(v)\geq W(v^{*}) = W((v^{*})^{T}) \geq W(B(0,t')).
\end{align*}
This completes the proof.

\end{document}